\documentclass[letterpaper,twocolumn,10pt]{article}
\usepackage{usenix-2020-09}

\usepackage{tikz}
\usepackage{amsmath}

\usepackage{diagbox}

\usepackage{filecontents}
\usepackage{tabularx}
\usepackage{booktabs}

\usepackage{booktabs}   
\usepackage{tabularx}  
\usepackage{array}      

\usepackage{amssymb}
\usepackage{multirow}
\usepackage{balance}  
\usepackage{flushend}

\usepackage{cite}
\usepackage{url}
\usepackage{xcolor} 
\usepackage{comment}
\usepackage{color}
\usepackage{algorithm}   
\usepackage{algorithmic}
\usepackage{xspace}
\usepackage{amsmath}
\usepackage{graphicx}
\usepackage{subcaption}
\usepackage{makecell} 
\usepackage{booktabs} 
\usepackage{enumitem}
\usepackage{pifont}
\usepackage{placeins}
\usepackage{amssymb}
\usepackage{multirow}
\usepackage[bottom]{footmisc}

\usepackage{caption}
\usepackage{stfloats}   
\usepackage{afterpage}

\allowdisplaybreaks

\newcommand{\pname}{{TARMM}\xspace}  
\newcommand{\phyan}[1]{\textbf{\textcolor{blue}{[#1]}}}

\usepackage{enumitem}
\setlist[itemize]{leftmargin=*}

\begin{document}

\date{}

  
\title{\pname: A Temporal Graph-Based 5G O-RAN System for Proactive Mobility Management in Delay-Critical Edge AI}
\title{\pname: A Temporal Graph-Based 5G Open RAN System for Delay-Critical Edge AI Offloading
}
\title{\pname: Temporal Graph-Based Mobility Management for Delay-Critical Edge AI Offloading in 5G Open RAN}
\title{\pname: Scaling Delay-Critical Edge AI Offloading in 5G O-RAN via Temporal Graph Mobility Management}
\author{
{\rm Peihao Yan}
\and
{\rm Yun Chen}
\and
{\rm Jie Lu}
\and
{\rm Qijun Wang}
\and
{\rm Huacheng Zeng}
\and
\\
\textnormal{ Department of Computer Science and Engineering, Michigan State University}
} 

\maketitle

\begin{abstract}
Emerging delay-critical edge AI applications, such as VR perception and real-time video analytics, impose stringent latency and reliability requirements on 5G networks. However, existing mobility management mechanisms are largely reactive and fail to adapt to dynamic network conditions, resulting in suboptimal handover decisions and degraded performance.
In this paper, we present \pname, a 5G Open Radio Access Network (O-RAN) system that optimizes user mobility management for delay-critical edge AI offloading. The core of \pname is a temporal graph model that captures the \textit{spatiotemporal} dynamics of the RAN across users and cells, enabling near real-time handover decisions. Building on this representation, we design a multi-agent reinforcement learning (MARL) framework with rule-based action masking and proactive resource preparation to ensure safe, stable, and efficient handovers. We implement \pname on a multi-cell indoor 5G O-RAN testbed and evaluate it using diverse VR workloads. Extensive experiments show that \pname reduces tail latency by up to 44\% and packet loss by up to 56\% compared to state-of-the-art approaches.
Source code and demo videos are available at: \url{https://margo-source.github.io/Margo/}

\end{abstract}

\vspace{-0.1in}
\section{Introduction}

Mobile devices such as smartphones, AR/VR headsets, and mobile robots are fundamentally constrained by limited battery capacity and on-device computational resources. As a result, advanced AI applications increasingly rely on edge AI offloading, where sensor data are transmitted to nearby edge GPU servers for inference, and results are returned in real time to support decision-making, rendering, and control. 
For delay-critical applications, communication latency is not merely a performance metric but a fundamental functional requirement. 
Applications such as immersive AR/VR and real-time robotics operate in closed-loop systems, where sensing, inference, and actuation must occur within strict latency bounds. Even short-lived latency spikes or jitter can disrupt user experience, degrade control stability, or lead to system-level failures.
Therefore, achieving low and stable communication delays is essential for edge AI offloading.

The communication delay in cellular networks is determined by many factors such as spectrum availability, radio link quality, cell load, and scheduling policies. Among these, handover plays a particularly critical role for mobile users. As a user equipment (UE) moves across cells, handover introduces transient disruptions in connectivity, resource reallocation, and scheduling, often resulting in delay spikes and increased jitter.
While handover decisions are made at the level of individual UEs, they are inherently influenced by global network conditions, such as target cell load and interference level. For example, selecting a target cell with strong signal strength but high congestion may increase queuing delay, whereas choosing a lightly loaded but distant cell may degrade radio link quality. The coupling between local mobility decisions and global network dynamics makes delay-aware handover management a challenging problem.

While handover management has been extensively studied under various objectives (e.g., maximizing throughput \cite{mehregan2025gcn}, improving load balancing \cite{chang2022decentralized,pateromichelakis2014graph}, enhancing fairness \cite{prado2023enabling}, and reducing signaling overhead \cite{shen2025decentralized}), 
most existing work relies on simplified network models or simulated datasets that are incapable of capturing the complexity and dynamics of real-world networks.
Moreover, most existing approaches were evaluated through numerical analysis, simulation, or network emulation, which often do not reflect the operational constraints and variability of deployed systems. 
The design and evaluation of handover management strategies in realistic network environments remain underexplored, particularly for delay-critical edge AI applications.

In this paper, we present \pname, a 5G Open Radio Access Network (O-RAN) system with optimized UE mobility management for delay-critical edge AI offloading.
\pname focuses on the handover optimization for cell-edge UEs, 
with the aim of jointly minimizing uplink and downlink communication delays while respecting UE's demands and cell load constraints.
Our key insight is that effective handover decisions must jointly consider temporal dynamics and spatial dependencies of the entire network. 
To this end, we design a handover framework that combines temporal graph representation learning, multi-agent reinforcement learning (MARL), and rule-based control mechanisms to determine \textit{when}, \textit{which} target cell, and \textit{how} to perform handover of individual UEs in dense small cell networks.



One core component of \pname is a temporal graph model. To effectively capture the complex interactions in multi-cell networks, we model the RAN as a temporal graph where both UEs and RUs are represented as nodes, and their time-varying interactions form dynamic edges. Each UE aggregates cross-layer information from neighboring cells, including radio measurements (SINR, RSRP, CQI), traffic statistics, and cell load conditions. A Temporal Graph Network (TGN) is employed to learn embeddings that encode both spatial dependencies (e.g., inter-cell interference and topology) and temporal dynamics (e.g., mobility patterns and traffic evolution). This representation enables the system to move beyond instantaneous measurements and instead reason over historical and contextual information, forming the foundation for predictive and proactive network control.

Based on the graph-based representations, we adopt a MARL model for UE handover decision making, where each UE is modeled as an autonomous agent that makes decisions based on local observations derived from temporal graph embeddings. While MARL provides strong adaptability, it learns through interaction with the real network, where exploration may lead to unsafe or suboptimal handover actions. For instance, an RL agent may temporarily hand over a UE to a distant cell for exploration, causing service disruption.
To address this challenge, we integrate rule-based constraints into the learning process. Domain knowledge (\textit{e.g.}, signal strength thresholds, connectivity requirements, and load constraints) is encoded as \textit{differentiable} action masks that filter out invalid or unsafe actions before handover execution. This integration ensures that all handover actions satisfy practical network constraints while allowing the learning model to optimize performance within the feasible action space.

We implement \pname on an indoor multi-cell 5G O-RAN testbed that comprises five cells and a number of smartphones, and evaluate its performance from both network and application perspectives. Experimental results show that \pname outperforms existing rule-based and state-of-the-art learning-based handover methods in terms of both latency and path loss rate.
From the network perspective, \pname reduces the 95th-percentile latency by up to 44\% and the path loss rate by up to 56\% compared to the state of the art. From the VR application perspective, \pname reduces the frame drop rate by up to 46\% and the frame freeze ratio by up to 61\% compared to the state of the art.




The contributions of this work are summarized as follows:

\begin{itemize}[leftmargin=0.15in,itemsep=0in,topsep=0in]
\item 
We propose a temporal graph-based representation of multi-cell 5G networks that captures both spatial interactions and temporal dynamics, enabling delay-aware decision making beyond instantaneous measurements.

\item 
We design a unified framework that integrates reinforcement learning with rule-based action masking, achieving both adaptability and safety in handover control.

\item 
We implemented and evaluated the proposed system on a realistic 5G O-RAN testbed, demonstrating substantial improvements in communication latency and stability for delay-critical edge AI applications.
\end{itemize}

To the best of our knowledge, this is the first work on system-level handover optimization for realistic cellular networks. It highlights the importance of integrating learning-based decision-making with domain knowledge and system-level design in next-generation RANs.

\vspace{-0.1in}
\section{Handover and Delays in 5G NR}

\vspace{-0.1in}
\subsection{5G Handover Protocols in 3GPP}


In 5G NR, handover is a network-controlled procedure specified in 3GPP \cite{3gpp38300, 3gpp38331, 3gpp38401}. 
The procedure consists of three phases. 

\begin{itemize}[itemsep=0in,topsep=0in]
\item 
\textit{Phase-1: Measurement and decision.}
Each UE continuously reports its radio link conditions to its home gNB, and the network determines whether to trigger handover.
In 5G NR, handover decisions are typically made based on the Event A3 criterion. Under this rule, handover is triggered if the signal power of a neighboring cell exceeds that of the serving cell by a hysteresis offset $\Delta_{\mathrm{A3}}$ for a time duration of $T_{\mathrm{TTT}}$.
This requirement is to filter out short-term fluctuations and reduce premature handovers.

\item 
\textit{Phase-2: Handover Preparation.} 
If handover is triggered, the serving gNB sends a Handover Request to the target gNB via the Xn or NG interface to transfer the UE context and QoS requirements and initiate resource preparation.
The target gNB then performs admission control and reserves the necessary resources for the incoming UE. 
If admitted, it replies with a Handover Request Acknowledge, which includes a dedicated Random Access Channel (RACH) preamble and a Radio Resource Control (RRC) container with the target cell configuration for subsequent access.

\item 
\textit{Phase-3: Handover Execution.} 
The UE performs RRC reconfiguration to detach from the serving cell and access the target cell. When triggered under weak RSRP conditions, unreliable synchronization and repeated random access, including multiple RACH attempts and retransmissions, can significantly prolong this phase, resulting in a transient service interruption where data transmission is suspended. 
During this interruption, uplink and downlink traffic continue to arrive but cannot be served, accumulating at the UE and DU buffers. After the UE reconnects to the target cell, the buffered data is released in bursts, causing temporary congestion at the MAC/RLC layers.
\end{itemize}



\begin{figure*}[t]
\centering
\includegraphics[ width=0.9\linewidth]{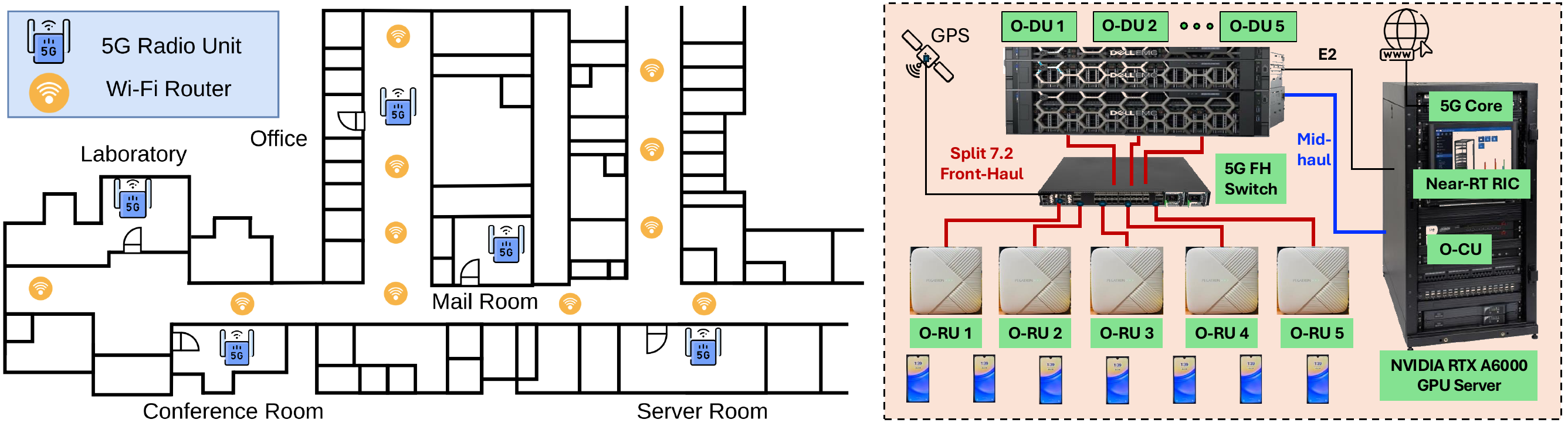}\vspace{0.05in}
\caption{An indoor 5G NR O-RAN testbed deployed in a university building.}\vspace{-0.1in}
\label{fig:testbed}
\end{figure*}

\vspace{-0.1in}
\subsection{Delay Measurements}

To understand the communication latency in incumbent networks, we conducted a measurement campaign in a university building equipped with 5G NR and Wi-Fi networks, as shown in Fig.~\ref{fig:testbed}.
Our indoor multi-cell 5G NR O-RAN testbed is built on the srsRAN project \cite{srsran_github}, which is fully compliant with 3GPP and O-RAN specifications.

\textbf{Multi-Cell Indoor 5G O-RAN.}\label{sec:indoor_5g_network}
The 5G O-RAN comprises five commercial radio units (RUs) that provide the coverage of the entire building floor, five distributed units (DUs), and one central unit (CU), Near-RT RIC, and Non-RT RIC.
The commercial RUs are from different vendors including Benetel and Pegatron. 
The DU and CU use the srsRAN protocol stacks \cite{srsran_github}. 
Near-RT and Non-RT RICs were deployed in a local NVIDIA A6000 GPU server. 
The system operates on the n78 band under an FCC experimental license.
The center frequency is 3350.01~MHz, and the subcarrier spacing is 30~kHz.
The bandwidth is 100~MHz and the MIMO configuration is 4x4. 
This network provides Internet access to 20+ smartphones, which are distributed within the building. 

The building is also equipped with a Wi-Fi network comprising dense Aruba Wi-Fi 6 routers \cite{aruba_wifi6_deployment}. 
The Wi-Fi network operates on the 5 GHz band with 40~MHz bandwidth. 
The routers use the same Extended Service Set Identifier (ESSID), providing building-wide coverage and enabling automatic roaming between routers.

\begin{figure}[t]
        \centering
        \begin{subfigure}[b]{0.51\linewidth} 
            \centering
            \includegraphics[trim=0 0 0 0, clip, width=1.04\linewidth]{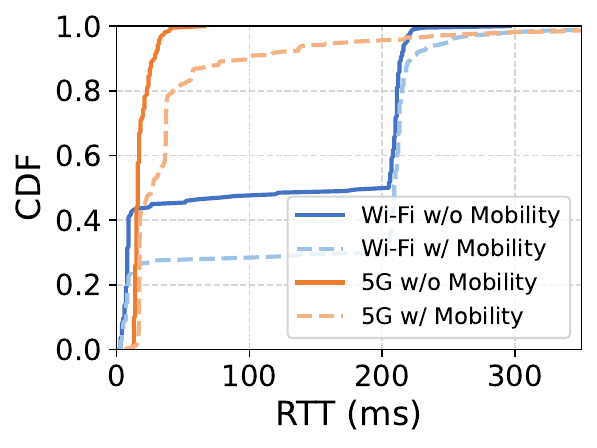}\vspace{-0.1in}
            \caption{CDF of measured RTT.} 
            \label{fig:wifi_5g_cdf}
        \end{subfigure}
        \hfill
        \begin{subfigure}[b]{0.48\linewidth}
            \centering
            \includegraphics[trim=0 -15 0 0, clip,width=1.05\linewidth]{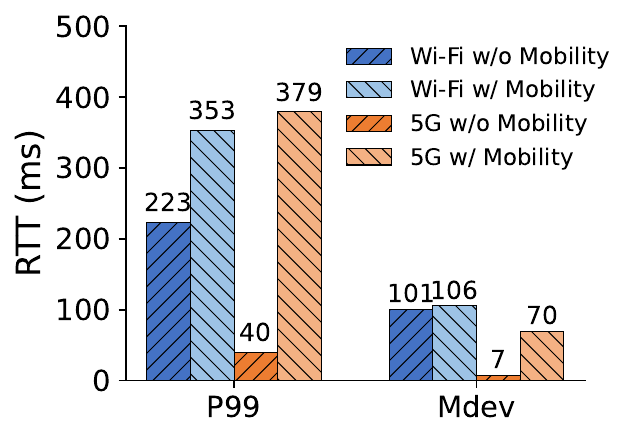}\vspace{-0.1in}
            \caption{99-percentile and std dev.}
            \label{fig:wifi_5g_p99}
        \end{subfigure}
        \caption{RTT measurement results of Wi-Fi and 5G.} \vspace{-0.1in}
        \label{fig:RTT_CDF_P99}
\end{figure}

\textbf{Measurement Results.}
We measure the round-trip time (RTT) of smartphones \texttt{ping} an edge GPU server under two cases: (i) the smartphones are stationary with no handover, and (ii) a smartphone moves across the entire floor. The experiments are conducted under very light traffic conditions for both 5G and Wi-Fi. 
Fig.~\ref{fig:wifi_5g_cdf} reports the results. Wi-Fi exhibits high latency and significant variability, with a large fraction of samples exceeding 200\,ms, primarily due to channel contention and backoff. In contrast, 5G achieves substantially lower and more concentrated RTT distributions, with the majority of data packets falling within 15–40\,ms even under mobility, owing to centralized scheduling.

A key observation for 5G is that UE mobility significantly impacts RTT performance, as shown in Fig.~\ref{fig:wifi_5g_cdf}. Fig.~\ref{fig:wifi_5g_p99} further compares the 99th-percentile latency and standard deviation with and without mobility. It confirms that mobility significantly increases both metrics.
The 99th-percentile RTT rises from 40\,ms to 124\,ms.

\begin{figure}
\centering
\includegraphics[trim=0 2 0 0, clip, width=\linewidth]{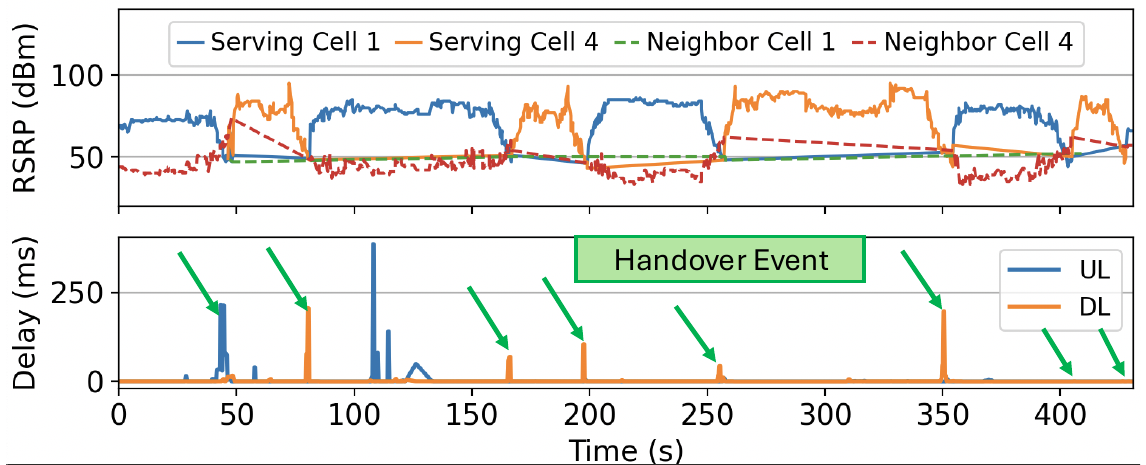}
\caption{Impact of handover events on queuing delay in 5G.}\vspace{-0.1in}
\label{fig:handover_event}
\end{figure}

\textbf{Queueing Delay During Handover.}
To examine handover behavior, we carry a Samsung smartphone generating 10~Mbps persistent \verb|iperf| downlink traffic and walk across two neighboring cells. 
The device performs automatic handover following the 3GPP Event A3 criterion. 
Fig.~\ref{fig:handover_event} shows the corresponding RSRP evolution together with the measured queueing delay. 
We observe that the queueing delay consistently exhibits sharp spikes aligned with handover events. 
Under steady connectivity, the delay remains below 20 ms, whereas during handover it can surge to over 200 ms. 
These delay spikes are short-lived but recurrent. 
Notably, the experiment is conducted under light cell load, indicating that the observed delay inflation is not due to resource saturation, but rather transient disruptions in buffering and scheduling during the handover procedure.

\begin{figure}
\centering
\includegraphics[trim=0 0 -10 0 , clip, width=0.93\linewidth]{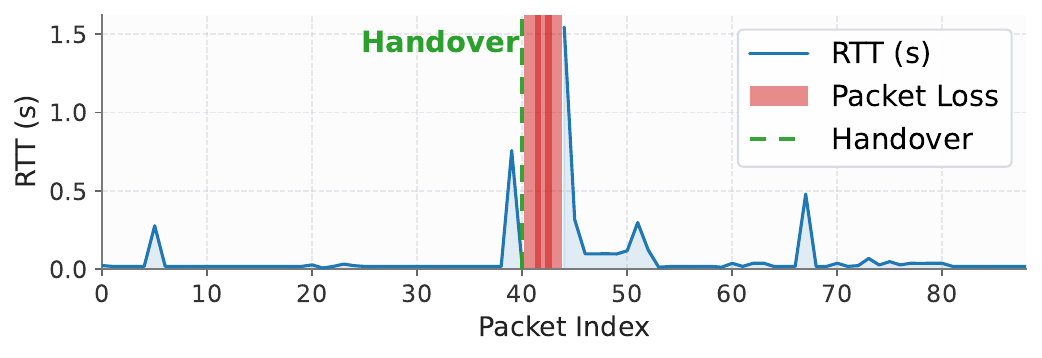}
\caption{Impact of handover on packet latency and reliability. }\vspace{-0.1in}
\label{fig:rtt_cell2cell3_zoom}
\end{figure}

\textbf{Packet Loss During Handover.}
Fig.~\ref{fig:rtt_cell2cell3_zoom} illustrates the impact of handover on latency and reliability. 
We observe that packet loss occurs within a short window immediately after the handover event (highlighted region around packets 40-45), indicating a transient service disruption during the transition. 
This period is also accompanied by a pronounced RTT spike exceeding 1.5~s, in sharp contrast to the baseline RTT, which remains consistently below 50~ms before and after handover. 
Outside the handover interval, RTT quickly returns to a stable low-latency regime with no packet loss. 
These results suggest that while handover is generally efficient, it can introduce brief but severe latency spikes and packet loss due to temporary interruption and buffering effects.


\begin{figure}
\centering
\includegraphics[trim=0 0 -25 0 , clip, width=\linewidth]{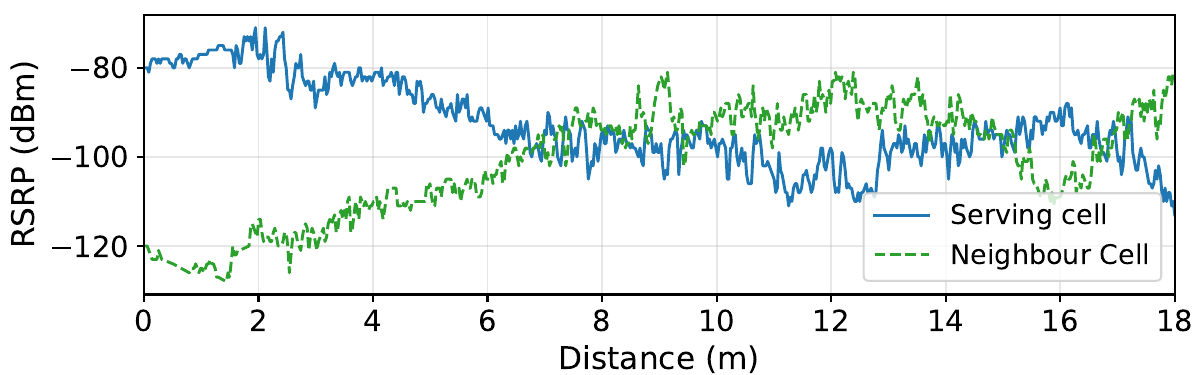}\vspace{0.05in}
\caption{Measured RSRP profile on a smartphone when it moves from one cell to another in our indoor 5G NR network.}\vspace{-0.1in}
\label{fig:RTT_RSRP_mobility}
\end{figure}

\vspace{-0.1in}
\subsection{Why Learning-Based Handover Needed?}

In current 5G systems, handover decisions are predominantly based on rule-based criteria such as Event A3, which triggers a handover when a neighboring cell’s signal strength exceeds that of the serving cell by a fixed threshold. While effective in macro-cellular networks, this approach becomes inadequate in micro- and small-cell deployments, particularly in urban and indoor environments. For example, indoor settings are characterized by rich multipath propagation, frequent occlusions, and complex layouts, leading to highly irregular and rapidly fluctuating signal strength even over short distances (see Fig.~\ref{fig:RTT_RSRP_mobility}). 
As a result, instantaneous signal measurements can be misleading, causing premature or unnecessary handovers, or delayed decisions that degrade link quality. This limitation is critical because handover events introduce transient service interruptions, which can lead to significant RTT spikes and packet loss even under light load conditions.

Moreover, RSRP-based handover strategies overlook other important factors such as cell load, queuing delay, and application-specific latency requirements, potentially resulting in suboptimal target cell selection and degraded performance. Since each network scenario exhibits unique spatial signal patterns and traffic dynamics, fixed threshold-based policies lack the adaptability required for reliable operation. Therefore, a learning-based handover solution is needed to capture complex spatiotemporal network characteristics, leverage historical and contextual information, and dynamically optimize handover decisions based on both radio conditions and overall network state.

\section{\pname: Design}
\vspace{-0.1in}







\begin{figure}[t]
\centering
\includegraphics[width=1.05\linewidth]{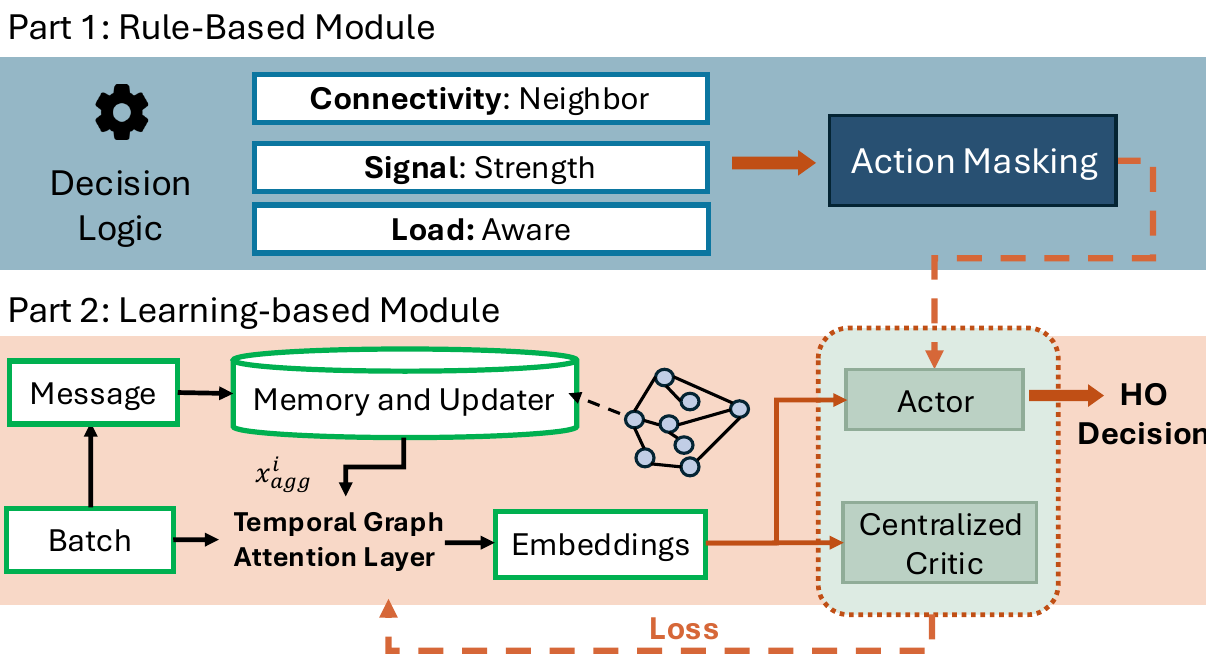}
\caption{The architecture of \pname.}
\label{fig:design}
\end{figure}

Fig.~\ref{fig:design} presents the high-level architecture of \pname, which consists of two key components: (i) temporal graph embedding and (ii) a graph-based actor–critic MARL framework with rule-based action masking. The learning-based module adapts handover decisions to dynamic network conditions, while the rule-based masking filters out unsafe actions using domain knowledge. We describe each component in detail below.

\vspace{-0.1in}
\subsection{Temporal Graph-Based Representation}

\textbf{Temporal Graph Modeling.}
The RAN environment under mobility exhibits both strong temporal and spatial dependencies. 
\textit{From temporal perspective}, user mobility and network conditions evolve continuously over time. The future state of a UE, including its serving cell, channel quality, and experienced delay, is highly correlated with its historical trajectory. Therefore, relying solely on instantaneous observations leads to myopic decisions that fail to anticipate upcoming handovers or delay variations. Capturing such temporal dependencies requires maintaining historical information of UE mobility and network dynamics.
\textit{From a spatial perspective}, the performance of each UE is inherently influenced by its surrounding environment, including neighboring cells and nearby UEs. In particular, neighboring cells determine candidate handover targets and their load conditions, while neighboring UEs compete for shared radio resources, affecting scheduling and delay performance. As a result, effective decision making requires modeling interactions among nodes in the network.

\begin{figure}[!t]
\centering
\includegraphics[trim=0 0 0 0 , clip, width=1\linewidth]{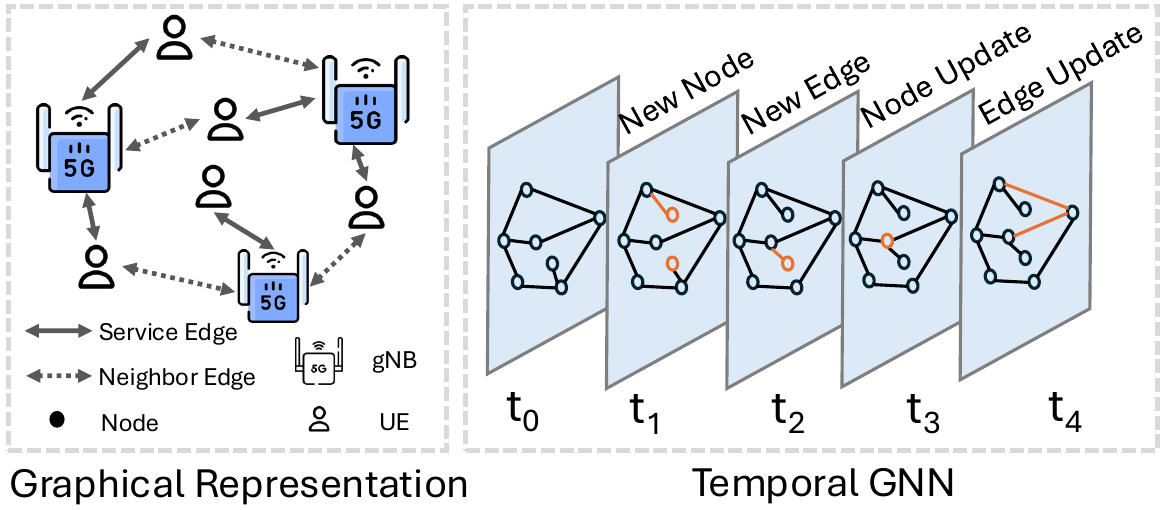}\vspace{0.1in}
\caption{Graph representation of an RAN.}\vspace{-0.15in}
\label{fig:graph_design}
\end{figure}

To jointly capture temporal evolution and spatial interactions, we model the RAN as a time-evolving graph, where nodes represent individual UEs and cells, and edges represent serving and neighboring links, as illustrated in Fig.~\ref{fig:graph_design}. Both node states and edge relationships are continuously updated to reflect mobility dynamics and inter-cell dependencies. Specifically, we adopt a Temporal Graph Network (TGN) to learn dynamic node representations by integrating historical information with neighborhood context. This formulation enables the model to capture both the temporal progression of UE states and the influence of neighboring cells and UEs, providing a unified framework for modeling spatiotemporal dependencies in the network.





\begin{figure}[t]
\centering
\includegraphics[trim=0 0 0 0 , clip, width=\linewidth]{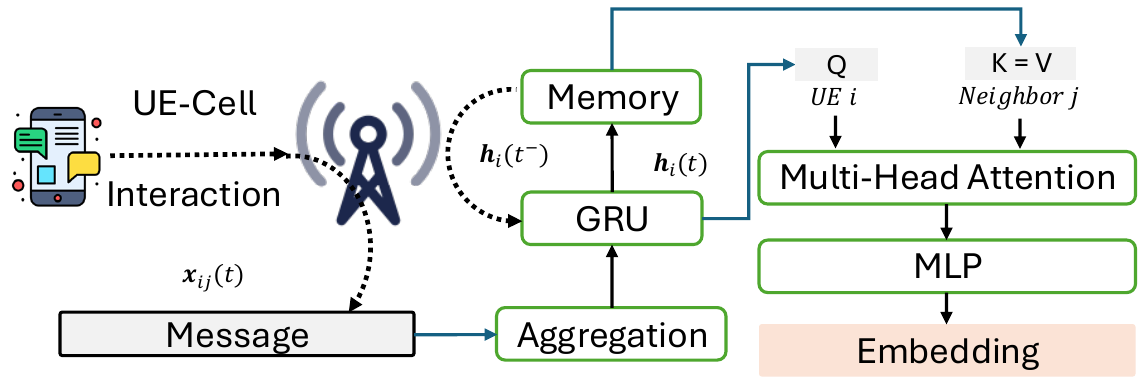}\vspace{0.05in}
\caption{ Temporal graph modeling and message passing in \pname.}\vspace{-0.1in}
\label{fig:Graph_design}
\end{figure}

\textbf{Message Passing and Feature Learning.}
We model the RAN system as a continuous-time dynamic graph represented by a sequence of time-stamped events, as shown in Fig.~\ref{fig:Graph_design}.
Let $\mathcal{G}(t) = (\mathcal{N}, \mathcal{E}(t))$ denote the graph, where $\mathcal{N}$ is the set of nodes corresponding to individual UE and base station (BS), and 
$\mathcal{E}(t)$ is the set of edges corresponding to each UE and its serving and neighboring BS. 
Each node maintains a time-evolving memory state $\mathbf{h}_i(t)$, encoding its historical observations.
The state of each node updates through message passing, triggered by the periodically KPM reportings from O-RAN's E2 interface.

For the graph, we define two types of events for message passing: 
(i)  \textit{edge event} and 
(ii) \textit{node event}. 

\begin{itemize}[itemsep=0in,topsep=0in]
\item 
\textit{An edge event} captures the interaction between a UE node $i$ and a BS node $j$, defined as $\mathbf{x}_{ij}(t) = (i, j, t, \mathbf{e}_{ij}(t))$, where $\mathbf{e}_{ij}(t)$ is constructed from the KPM in Table~\ref{tab:inputdata}. 
This event triggers message generation for both UE node $i$ and BS node $j$.
The generated messages are:
\begin{align}
\mathbf{m}_i(t) &= 
\Big[
\mathbf{h}_i(t^-),\;
\mathbf{h}_j(t^-),\;
\Delta t_i,\;
\mathbf{x}_{ij}(t)
\Big], \\
\mathbf{m}_j(t) &= 
\Big[
\mathbf{h}_j(t^-),\;
\mathbf{h}_i(t^-),\;
\Delta t_v,\;
\mathbf{x}_{ij}(t)
\Big],
\end{align}
where $\mathbf{h}_i(t^-)$ denotes the memory state of node $i$ immediately before time $t$, and $\Delta t_i$ is the elapsed time since the last event of node $i$.

\item 
\textit{A node event} is designed for node-specific state updates, defined as $\mathbf{x}_i(t) = (i, t, \mathbf{n}_i(t))$, where $\mathbf{n}_i(t)$ is constructed based on the KPM in Table~\ref{tab:inputdata} that are associated with this UE or BS node. 
The generated message can be written as follows:
\begin{equation}
\mathbf{m}_i(t)
=
\Big[
\mathbf{h}_i(t^-),\;
\Delta t_i,\;
\mathbf{x}_i(t)
\Big].
\end{equation}
\end{itemize}

For a UE or BS node in the graph, it may receive multiple messages within a training batch.
In this case, messages are aggregated using a most-recent aggregator, which selects the message with the latest timestamp to form the aggregated message $\bar{\mathbf{m}}_i(t)$.
When a node receives a message, its memory is updated via a gated recurrent unit (GRU):
\begin{equation}
\mathbf{h}_i(t) = \mathrm{GRU}\big(\bar{\mathbf{m}}_i(t), \; \mathbf{h}_i(t^-)\big).
\end{equation}
The GRU enables each UE and cell to capture long-term temporal dependencies, such as persistent congestion or gradual signal degradation, which are difficult to infer from instantaneous KPM.

\begin{table}[t]
\fontsize{6.5}{7}\selectfont
\caption{KPM data used for message representation in graph.}
\label{tab:inputdata}
\centering
\setlength{\tabcolsep}{3pt}
\begin{tabular}{@{}p{1.1cm}p{2.8cm}>{\centering\arraybackslash}p{0.6cm}p{3.6cm}@{}}
\toprule
\textbf{Source} & \textbf{Feature} & \textbf{UL/DL} & \textbf{Description} \\
\midrule
\multirow{3}{*}{UE-BS Edge}
& Serving RSRP / RSRQ / SINR & DL & Measurements from serving cell. \\
& Neighbor RSRP/ RSRQ/ SINR & DL & Measurements from neighboring cells. \\
& RAN / AMF / UE / DU IDs & -- & Cross-layer identifiers. \\
\midrule
\multirow{8}{*}{UE-Node}
& Throughput & DL/UL & Per-UE achieved data rate. \\
& Queuing Delay & DL/UL & Packet delay at the RLC layer. \\
& Air Interface Delay & UL & Transmission delay over the air. \\
& Packet Drop Rate & DL & Fraction of dropped packets. \\
& PRB Usage & DL/UL & Number of PRBs allocated to the UE. \\
& Traffic Volume & DL/UL & Transmitted data volume. \\
& CQI / RSRP / SINR & UL & Uplink radio quality indicators. \\
& UE Identifier & -- & UE identifiers for cross-layer. \\
\midrule
\multirow{3}{*}{Cell-Node}
& Cell Throughput & DL/UL & Aggregated throughput at the cell level. \\
& PRB Usage & DL/UL & Total PRBs used in the cell. \\
& PRB Utilization & DL/UL & Percentage of PRB utilization. \\

\bottomrule
\end{tabular}
\vspace{-0.1in}
\end{table}

\textbf{Attention-Based Embedding.}
To mitigate memory staleness and incorporate neighborhood dynamics, we compute node $i$'s embedding $\mathbf{z}_i(t)$ using a temporal multi-head attention mechanism over node $i$'s recent temporal neighborhood:
\begin{equation}
\mathbf{z}_i(t) = \mathrm{MLP}\Big( \mathbf{h}_i(t) \;\Vert\; \tilde{\mathbf{h}}_i(t) \Big),
\label{eq:graph_output}
\end{equation}
where 
$\Vert$ is the operator of vector concatenation
and
$\tilde{\mathbf{h}}_i(t) = \mathrm{MultiHeadAttn}\big( \mathbf{q}_i(t), \mathbf{K}_i(t), \mathbf{V}_i(t) \big)$.
In this attention, the query is defined as the current node memory,
i.e.,
$\mathbf{q}_i(t) = \mathbf{h}_i(t)$,
and the keys and values are constructed from its temporal neighbors:
\begin{equation}
\mathbf{K}_i(t) = \mathbf{V}_i(t) =
\Big[
\mathbf{h}_j(t_j^-) \;\Vert\; \mathbf{x}_{ij}(t_j) \;\Vert\; \phi(t - t_j)
\Big]_{j \in \mathcal{N}_i(t)},
\end{equation}
where $\mathcal{N}_i(t)$ denotes the set of neighbors of node $i$, and $t_j$ is the timestamp of the most recent interaction between nodes $i$ and $j$ prior to time $t$. 
$\mathbf{h}_j(t_j^-)$ is the memory of node $j$ immediately before time $t_j$. $\phi(\cdot)$ is a time encoding function capturing recency effects.
This attention mechanism enables the model to selectively focus on neighboring cells whose recent interactions are most relevant to the UE’s current radio and QoE conditions.






\vspace{-0.1in}
\subsection{Graph-Based Actor-Critic MARL}

With the node embedding $\mathbf{z}_i(t)$, we formulate the handover problem as a multi-agent decision-making process, where each UE $i \in \mathcal{U}(t)$ is treated as an agent. Each agent makes handover decisions based on its local observation, while sharing a common policy across all agents. 
The observation is defined as $\mathbf{o}_i(t) = \mathbf{z}_i(t)$. The policy is parameterized as $\pi_{\theta}\big(\mathbf{a}_i(t)\mid \mathbf{o}_i(t)\big)$, where $\mathbf{a}_i(t)$ denotes the action of agent $i$ at time $t$, and $\theta$ represents the shared policy parameters.



\textbf{Action and Observation Space.}
At each time step $t$, each UE agent $i \in \mathcal{U}(t)$ observes a local observation $\mathbf{o}_i(t)$ derived from the temporal graph embedding $\mathbf{g}_i(t)$.
This observation has already encoded: (i) UE channel quality, (ii) neighboring cell conditions, and (iii) temporal dynamics reflected through past interactions.
Each UE agent selects a handover action defined as
\begin{align}
\mathbf{a}_i(t) = \big(p_i^{\mathrm{HO}}(t),\, c_i^{\mathrm{target}}(t)\big),
\end{align}
where $p_i^{\mathrm{HO}}(t) \in \{0,1\}$ is  handover decision and $c_i^{\mathrm{target}}(t)$ specifies the target cell index.





\begin{figure}[t]
\centering
\includegraphics[trim=0 0 0 0 , clip, width=\linewidth]{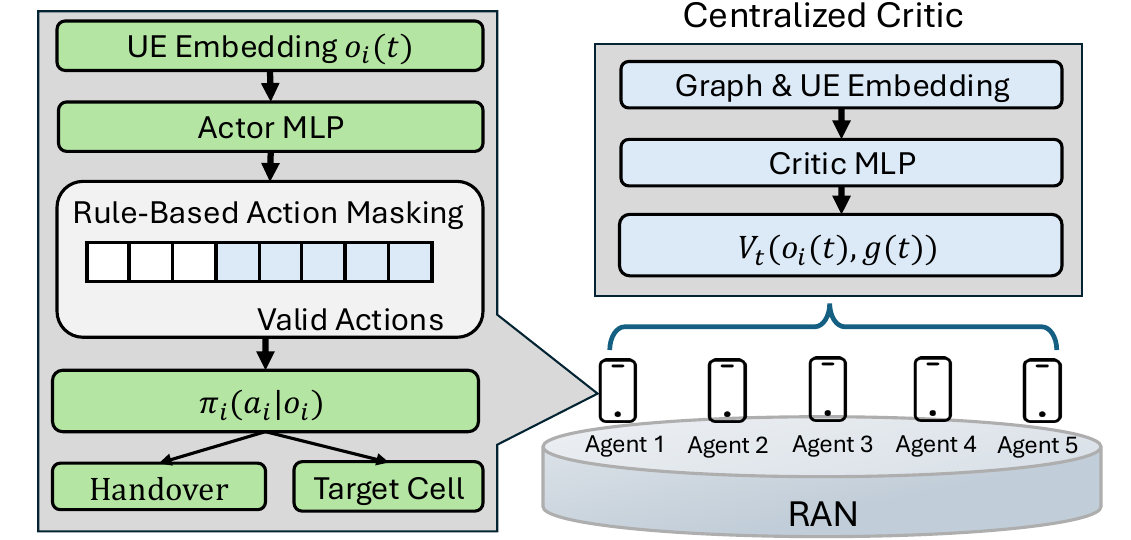} \vspace{0in}
\caption{Diagram of proposed actor-critic MARL framework.}\vspace{-0.1in}
\label{fig:AC_design}
\end{figure}

\textbf{Learning Objective.}
A handover event triggers network-side RRC reconfiguration for the UE, introducing a transient service interruption that manifests as elevated queuing delay at the O-RAN DU.
For each UE agent $i \in \mathcal{U}(t)$, let $D_i$ denote its delay requirement, and let $\tau_i(t)$ denote the experienced packet delay at time $t$. 
We define $\tau_i(t)$ as the sum of uplink and downlink queuing delays, i.e.,
$\tau_i(t) = \tau_i^{\mathrm{UL}}(t) + \tau_i^{\mathrm{DL}}(t)$.
If $\tau_i(t) \le D_i$, the delay requirement is satisfied and no penalty is incurred. Otherwise, we quantify the violation as a normalized delay deficit. The delay regret of UE $i$ at time $t$ is defined as:
\begin{align}
J_i(t) = \max\left( \frac{\tau_i(t) - D_i}{D_i}, \, 0 \right).
\end{align}

This regret captures the end-to-end latency of a UE. 
In particular, during handover, packets are temporarily buffered due to link interruption, scheduling delay, and resource reallocation, leading to queue buildup at the RLC/MAC layers.
The queuing delay reflects both the instantaneous channel condition and the stability of the connection. 
Therefore, minimizing $J_i(t)$ enables the agent to take the actions that can reduce handover-induced latency spikes and maintain delay-critical service requirements.
We define the reward of agent $i$ as the negative of its delay regret:
$R_i(t) = - J_i(t)$.

\textbf{Actor-Critic Network.}
We adopt a parameter-sharing multi-agent actor--critic architecture with decentralized actors and a centralized critic, as illustrated in Fig.~\ref{fig:AC_design}. Each UE acts as an agent and makes decisions based on its local observation $\mathbf{o}_i(t)$.
The actor takes the UE embedding $\mathbf{o}_i(t)$ as input and feeds it into an MLP to produce a policy $\pi_i(a_i \mid \mathbf{o}_i)$. 
A rule-based action masking module is applied before sampling to filter out infeasible actions, and the policy is renormalized over the valid action space. 
The final action is multi-discrete, consisting of handover triggering and target cell selection.
The actor outputs a multi-discrete action with two categorical heads for handover triggering and target cell selection. 
The centralized critic estimates the state value using both local and global context. 
The centralized critic leverages both local and global context to estimate the state value. 
Specifically, it takes as input the concatenation of the UE embedding 
$\mathbf{o}_i(t)$ and a graph embedding $\mathbf{g}(t)$, where $\mathbf{g}(t)$ is obtained via mean pooling over all node embeddings, and outputs the value function $V_t(\mathbf{o}_i(t), \mathbf{g}(t))$.

\vspace{-0.1in}
\subsection{Rule-Based Action Masking}





In MARL for handover decision-making in O-RAN, the learned policy is inherently stochastic, requiring a balance between exploration and exploitation. While exploration is essential for discovering optimal strategies, it can also lead to unsafe or impractical actions during training and deployment. For instance, a UE may be driven to connect to a far-away RU despite having a strong nearby serving cell, or be redirected from an underutilized RU to a heavily congested one purely for exploration. Such actions can cause severe performance degradation, including  increased latency and service interruption. Therefore, unconstrained exploration is undesirable in this practical networks.

To ensure safe and stable handover decisions, we incorporate rule-based action masking into the MARL policy. 
Specifically, we define a binary mask $\mathbf{M}_i(t) \in \{0,1\}^{|\mathcal{A}_i|}$ over the action space $\mathcal{A}_i$, where each entry indicates whether a candidate action is valid. The masked policy is given by:
\begin{align}
\tilde{\pi}_{\theta}(\mathbf{a}_i(t)\mid \mathbf{o}_i(t))
=
\frac{
\pi_{\theta}(\mathbf{a}_i(t)\mid \mathbf{o}_i(t)) \cdot \mathbf{M}_i(\mathbf{a}_i(t), t)
}{
\sum_{\mathbf{a}' \in \mathcal{A}_i}
\pi_{\theta}(\mathbf{a}'\mid \mathbf{o}_i(t)) \cdot \mathbf{M}_i(\mathbf{a}', t)
},
\label{eq:masked_grad}
\end{align}
which renormalizes the probability distribution over valid actions only.

\textbf{Domain Knowledge–Based Masking.}
To ensure that the learned policy produces valid and deployable handover decisions, we introduce three action-masking constraints derived from domain knowledge of 5G network operations.

\begin{itemize}[noitemsep, topsep=0pt]

\item 
\textit{Connectivity masking.}
A target cell must be observable in the UE's measurement reports. Let $\mathcal{N}'_i(t)$ denote the set of neighboring cells reported by UE $i$. Then, the connectivity mask is defined as:
\begin{align}
M^{(1)}_{i,j}(t) = \mathbb{I}\big(j \in \mathcal{N}'_i(t)\big),
\end{align}
where $\mathbb{I}(\cdot)$ return 1 if condition is true and 0 otherwise.


\item 
\textit{Signal strength masking.}
Although the MARL policy incorporates the target cell's signal strength alongside other KPMs, it may be underweighted relative to other KPM inputs, potentially leading to handover decisions toward weak cells. Insights from our experiments indicate that a candidate cell must provide sufficient signal strength to ensure reliable handover execution. Let $P_i^{j}(t)$ denote the RSRP of UE $i$ from candidate cell $j$, and let $\gamma$ denote the minimum RSRP threshold for handover. 
Then, the signal-based mask is defined as:
\begin{align}
M^{(2)}_{i,j}(t) = \mathbb{I}\big(P_i^{j}(t) \geq \gamma\big).
\end{align}

\item
\textit{Load-aware masking.}
In practice, a UE should not be handed over to a highly congested cell.
Thus, we restrict the action space based on cell load. We define the load of cell $j$ as $\rho_j(t) = \frac{1}{B_j} \sum_{i \in \mathcal{U}_j(t)} b_{i,j}(t)$, where $\mathcal{U}_j(t)$ is the set of UEs associated with cell $j$. Let $\eta_j$ denote the load threshold for cell $j$. The load-aware mask is then defined as:
\begin{align}
M^{(3)}_{i,j}(t) = \mathbb{I}\big(\rho_j(t) \leq \eta_j\big),
\end{align}
which filters out candidate cells whose load exceeds the allowable threshold.

\end{itemize}

Combining these three cases, the overall mask applied to the MARL is:
\begin{align}
M_{i,j}(t) = M^{(1)}_{i,j}(t) \cdot M^{(2)}_{i,j}(t) \cdot M^{(3)}_{i,j}(t),
\end{align}
which ensures that only feasible, high-quality, and non-congested target cells are considered during action selection.

\textbf{Model Differentiability.}
Differentiability is critical for efficient learning. Since the mask $\mathbf{M}_i(\mathbf{a}, t)$ is deterministic and independent of the policy parameters $\theta$, gradients propagate only through the original policy $\pi_\theta$ and do not alter the underlying learning dynamics. Taking the gradient of the masked log-policy yields:
\begin{align}
\nabla_\theta \log \tilde{\pi}_\theta
=
\nabla_\theta \log \pi_\theta
-
\nabla_\theta \log
\sum_{\mathbf{a}' \in \mathcal{A}_i}
\pi_\theta(\mathbf{a}'\mid \mathbf{o}_i(t)) \cdot \mathbf{M}_i(\mathbf{a}', t),
\end{align}
which corresponds to a reweighted version of the original policy gradient over the valid action space.

\vspace{-0.1in}
\subsection{Online Training Process}

\pname includes three networks:
graph encoder network parameterized by $\psi$, MARL actor network parameterized by $\theta$, and MARL critic network parameterized by $\phi$.
The actor-critic networks are updated using standard PPO loss function, and the graph encoder network is updated by both actor loss and self-supervised link prediction loss. 
Specifically, they are updated as follows:
\begin{align}
(\theta, \psi) &\leftarrow 
(\theta, \psi) - \alpha_\pi \nabla_{\theta, \psi} \mathcal{L}_{\text{actor}}, \nonumber \\
(\phi, \psi)         &\leftarrow (\phi, \psi) - \alpha_v \nabla_{\phi, \psi} \mathcal{L}_{\text{critic}}, \label{eq:loss}\\
\psi         &\leftarrow \psi - \alpha_{\text{TGN}} \nabla_\psi \mathcal{L}_{\text{TGN}},\nonumber 
\end{align}
where 
$\alpha_\pi$, $\alpha_v$, and $\alpha_{\text{TGN}}$ are the learning rates 
of actor, critic, and graph updates.
$\mathcal{L}_{\text{actor}}$ and $\mathcal{L}_{\text{critic}}$ are actor-critic PPO losses (see Appendix~\ref{app:ppo}).

In particular, $\mathcal{L}_{\text{TGN}}$ in Eq.~\eqref{eq:loss} is adopted to stabilize representation graph learning.
It is defined as a self-supervised link prediction loss:
\begin{align}
\mathcal{L}_{\text{TGN}}
=
-\frac{1}{N}\sum_{(i,j)\in\mathcal{D}}
\Big[
y_{ij}\log \sigma(\hat{y}_{ij})
+
(1-y_{ij})\log\big(1-\sigma(\hat{y}_{ij})\big)
\Big],
\end{align}
where $\mathcal{D}$ is the set of sampled UE-RU pairs and $N = |\mathcal{D}|$. 
$y_{ij} \in \{0,1\}$ indicates whether the UE-cell interaction is observed, $\hat{y}_{ij} = \mathbf{z}_i^\top \mathbf{z}_j$ is the predicted link score computed from node embeddings and $\sigma(\cdot)$ is the sigmoid function.
To validate the quality of learned representations, we further analyze the temporal dynamics and consistency of TGN embeddings in Appendix~\ref{app:embedding}, where we show that the embeddings exhibit smooth temporal evolution and strong locality in representation space.

\begin{figure} [t]
\centering
\includegraphics[trim=0 0 0 0 , clip, width=\linewidth]{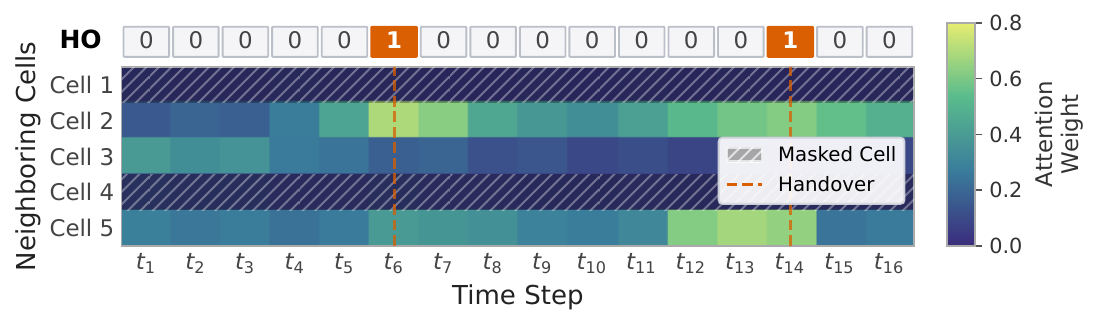}
\caption{Illustration of handover decision-making process under action masking.}\vspace{-0.1in}
\label{fig:Rloutput_heatmap}
\end{figure}

\vspace{-0.1in}
\subsection{Proactive Resource Preparation}

Fig.~\ref{fig:Rloutput_heatmap} illustrates the sequential handover decisions of the RL agent over time. The top row shows the binary handover decisions, while the heatmap below visualizes action preferences over neighboring cells. Hatched entries indicate masked cells removed from the feasible action space by rule-based constraints, restricting the agent to valid candidates only. The results show that the policy consistently concentrates on a specific feasible target cell with high preference, indicating that the RL agent learns to jointly decide \textit{when} and \textit{where} to hand over while respecting network constraints.

\textbf{Proactive Resource Preparation.}
While optimized handover decisions can reduce connectivity disruption, they do not address transient delay spikes immediately after handover. During this phase, the target DU treats the UE as a new arrival, resulting in cold-start scheduling, temporary queue buildup, and increased latency.

\begin{figure}[t]
\centering
\includegraphics[trim=0 0 0 0 , clip, width=\linewidth]{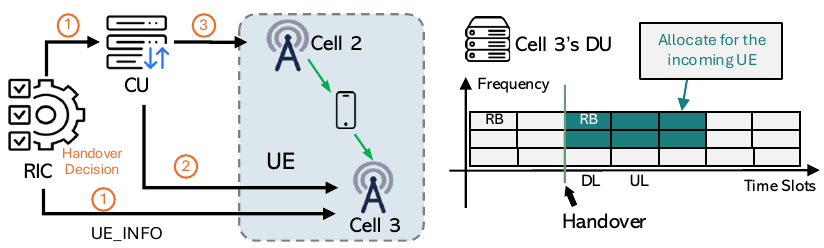}\vspace{0.1in}
\caption{Illustration of proactive resource reservation at the target cell.}\vspace{-0.1in}
\label{fig:proactive}
\end{figure}

To address this issue, we introduce a simple yet effective proactive resource reservation mechanism at the target DU. As shown in Fig.~\ref{fig:proactive}, once the RIC makes a handover decision, it sends the UE’s state information (e.g., recent resource usage and QoS requirements) to the target cell before the handover is executed. The target DU then reserves resources in advance, with the reserved amount defined as $b_i^{\mathrm{targ}} = \kappa \cdot b_i^{\mathrm{serv}}$, where $b_i^{\mathrm{targ}}$ and $b_i^{\mathrm{serv}}$ denote the PRBs at the target and serving cells, respectively, and $\kappa \in (0,1]$ is a scaling factor (set to 1 when the target cell is not overloaded). The UE is also assigned the highest scheduling priority upon arrival to ensure immediate service.

This lightweight mechanism significantly reduces post-handover queuing and scheduling delay. By enabling pre-allocation and informed scheduling, the UE can be served promptly after attachment, mitigating cold-start effects and stabilizing latency. More importantly, the approach is fully compatible with 3GPP and O-RAN, requiring no changes to standard handover procedures or scheduler design.

\begin{figure}[t]
        \centering
        \begin{subfigure}[b]{0.495\linewidth} 
            \centering
            \includegraphics[trim=0 0 0 0, clip, width=1.05\linewidth]{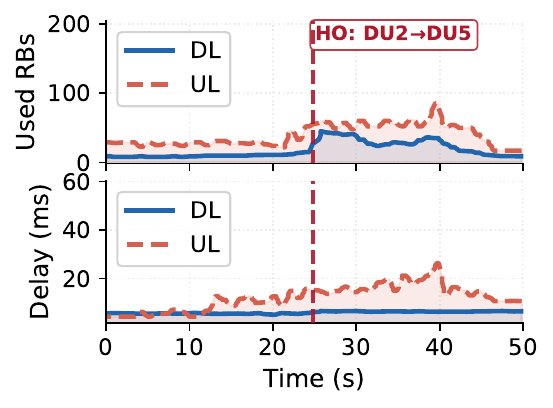}\vspace{-0.1in}
            \caption{w/ resource reservation.} 
            \label{fig:With_Proactive}
        \end{subfigure}
        \hfill
        \begin{subfigure}[b]{0.495\linewidth}
            \centering
            \includegraphics[trim=0 0 0 0, clip,width=1.05\linewidth]{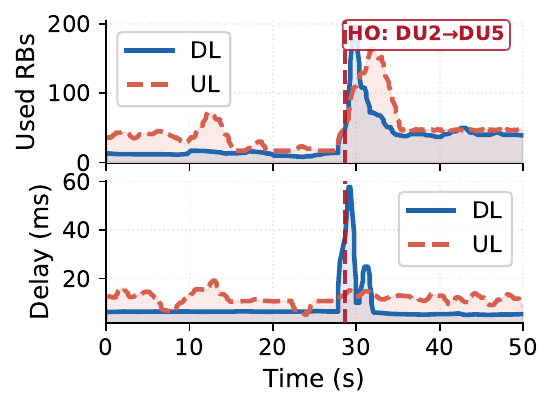}\vspace{-0.1in}
            \caption{w/o resource reservation.}
            \label{fig:Without_Proactive}
        \end{subfigure}
        \caption{Comparison of network resource utilization and UE's delay between two cases during handover events.} \vspace{-0.1in}
        \label{fig:Proactive_test}
\end{figure}

\textbf{Effectiveness of Proactive Handover.}
We evaluate this approach using a mobile UE with persistent 10\,Mbps uplink and 30\,Mbps downlink traffic. Fig.~\ref{fig:Proactive_test} compares performance with and without proactive resource reservation. With proactive preparation, the UE experiences a smooth handover transition, with gradual resource adaptation and consistently low delay. In contrast, without reservation, the target cell reacts only after handover completion, leading to a sudden surge in uplink demand and a sharp delay spike (up to 60\,ms) due to transient buffering and scheduling contention.

\vspace{-0.1in}
\section{Implementation}
\label{sec:implementation}


\vspace{-0.1in}
To emulate realistic network conditions, a set of smartphones remains stationary to generate background traffic and maintain cell load, while another set moves along predefined routes that traverse multiple cells to trigger handovers. 
These routes cover the entire network area. 
The measurement data is collected while continuously traversing the central regions of cells, focusing on mobility periods where handovers occur, with handover events observed approximately every 5-10 seconds.

\textbf{Near-RT RIC and KPM Details.}
All learning-based components are implemented in the O-RAN Software Community (OSC) Near-RT RIC \cite{oran_ric}, 
The RIC connects to one CU and five DUs via the E2 interface. UEs report their radio measurements (e.g., signal quality indicators) to the network every 120\,ms, while fine-grained network statistics are collected from the CU and DUs at a 10\,ms interval through E2 subscriptions. 

\textbf{Graph Embedding and MARL Parameters.}
We use a temporal graph to model dynamic UE-RU interactions. 
Each node maintains a 16-dimensional memory, updated via incoming messages and an aggregation function. 
For each event, we perform one round of temporal message passing followed by a TransformerConv layer (2 heads, dropout 0.1), projecting node embeddings from 16 to 32 dimensions. 
Relative time encodings are used as edge attributes, and node-type embeddings distinguish between UE and DU nodes.
The encoder outputs 32-dimensional UE embeddings $\mathbf{z}_i(t)$ and a graph-level embedding $\mathbf{g}(t)$ via mean pooling.
On top of the graph encoder, we adopt a parameter-sharing actor-critic architecture. 
The actor takes $\mathbf{z}_i(t)$ as input and feeds it into an MLP with hidden size (64, 64) (ReLU + LayerNorm) to generate multi-discrete actions. 
The critic takes $[\mathbf{z}_i(t), \mathbf{g}(t)]$ as input and uses an MLP (64, 64) with a linear head to estimate $V_{\phi}\big(\mathbf{z}_i(t), \mathbf{g}(t)\big)$.






\vspace{-0.1in}
\section{Experimental Evaluation}
\label{sec:evaluation}
\vspace{-0.1in}
We would like to answer the following questions through the evaluation.
\begin{itemize}[noitemsep, topsep=0pt]
\item 

\textbf{Q1:}
How does \pname impact RTT and packet loss for smartphones under varying network loads?
(\S\ref{main_result})


\item 
\textbf{Q2:}
How does \pname compare with state-of-the-art mobility management solutions? 
(\S\ref{subsection:State-of-the-Art})

\item 
\textbf{Q3:}
What is the impact of key design components in \pname, including Temporal Graph modeling and rule-based action masking?
(\S\ref{subsec:Ablation})

\item 
\textbf{Q4:}
How does \pname scale with increasing numbers of users and cells?
(\S\ref{subsec:Scalability})

\item 
\textbf{Q5:}
How does \pname improve user experience for VR applications under diverse AI workloads?
(\S\ref{VR_experiments})

\end{itemize}






\vspace{-0.1in}
\subsection{Performance Metrics and Baselines}

In our experiments, we use the following metrics to quantify the performance of \pname: 
uplink queuing delay, downlink queuing delay, packet loss rate, and the defined reward function.
We also compare \pname against the following SOTA learning-based baselines. 
\begin{itemize}[noitemsep, topsep=0pt]
\item 
\textbf{3GPP Standard~\cite{3gpp_sp99100}.}
The standard Event A3-based handover mechanism defined in 3GPP specifications, where a handover is triggered when the RSRP of a neighboring cell exceeds that of the serving cell by a predefined offset for a time-to-trigger duration.

\item
\textbf{MR-PHO \cite{hassan2022vivisecting}.}
MR-PHO stands for measurement-report-based predictive handover.
It is a learning based handover approach, leveraging historical measurement reports to predict future signal conditions for proactively triggering handovers.

\item
\textbf{Adaptive Parameter Tuning (APT) \cite{alsuhli2021mobility}.}
It employs a deep reinforcement learning for mobility load management, where each cell dynamically adjusts handover bias parameters to influence UE association decisions.
\end{itemize}


\vspace{-0.1in}
\subsection{Main Results}
With \pname deployed in our live 5G O-RAN testbed, we first conduct a comprehensive evaluation from a UE's perspective. 
Our data collection campaigm involves four Samsung smartphones, along with many other smartphones to generate background traffic for the control of cell traffic load. 
Of those four smartphones, one is stationary, placed close to an RU. 
This smartphone does not have mobility.
It is used as a reference serving as the upper bound for mobility management. 
The other three smartphones are carried by the same person, walking around the entire network coverage area. 
The mobility of these three smartphones are managed by three different handover control algorithms:
one uses in-house 3GPP handover algorithm;
one uses \pname; 
and 
one does not have explicit mobility management (No-MM).
In the case of No-MM, handover is disabled at the CU for this smartphone, forcing it to reconnect only after link failure when crossing cell boundaries.

During the experiments, the RTT and packet loss rate of those four smartphones are continuously recorded for data analysis. 
Apparently, the performance of a UE depends on the traffic load of the network. 
Therefore, we define two network traffic load cases using other smartphones: 
(i) light traffic load (10\% to 20\% of cell capacity), 
and
(ii) heavy traffic load (about 80\% of cell capacity). 
Apparently, the performance is also dependent on the traffic type, packet size of the UE under test. 
Therefore, we design two experiments: 
(i) \texttt{ping} small-sized bursty data traffic and (ii) \texttt{iperf} persistent data traffic.

\label{main_result}
\begin{figure}[t]
    \centering

    \begin{subfigure}[b]{\linewidth}
        \centering
        \includegraphics[width=0.54\linewidth]{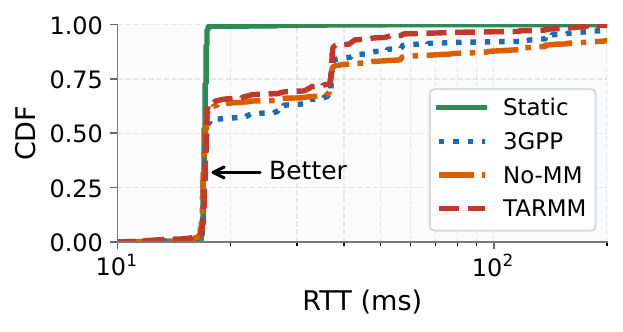}
        \includegraphics[trim=5 -15 8 0, clip, width=0.44\linewidth]{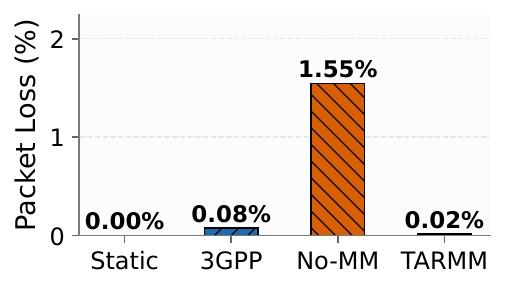}
        \vspace{-0.075in}
        \caption{All cells have light traffic loads.}
        \label{fig:light_ping_rtt1}
    \end{subfigure}
    \hfill
    \begin{subfigure}[b]{\linewidth}
        \centering
        \includegraphics[width=0.54\linewidth]{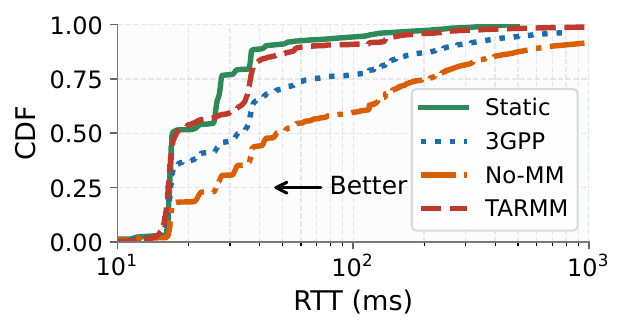}
        \includegraphics[trim=5 -15 8 0, clip, width=0.44\linewidth]{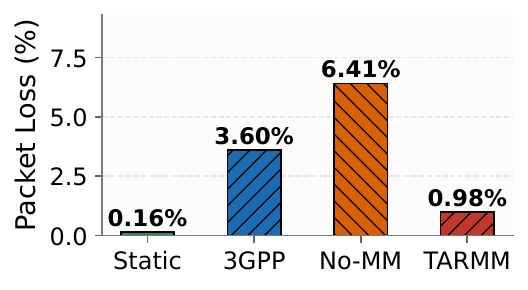}\vspace{-0.075in}
        
        \caption{All cells have heavy traffic loads.}
        \label{fig:light_ping_rtt2}
    \end{subfigure}\vspace{0.075in}    
    \caption{Comprehensive measurement results of RTT and packet loss rate when  UEs \texttt{ping} an edge GPU server.
    }\vspace{-0.1in}
    \label{fig:ping_RTTandLoss}
\end{figure}

\textbf{Ping Data Traffic on UE.} \ \ 
The \texttt{ping} command on smartphones generates very small data packets, representing the highly bursty traffic from the end users. 
In total, we collected 5 hours of \texttt{ping} data traces from those four smartphones over 3 days. 
Fig.~\ref{fig:ping_RTTandLoss} reports our experimental results. 
We have the following observations.
\textbf{First}, static UE has the best performance in terms of both RTT and PER. 
When the cell has light traffic load, most of its RTTs remain with 20~ms, and its PLR stay on zero.
This is as expected, thanks to its strong signal and the network's centralized scheduling mechanism. 
\textbf{Second}, the mobility significantly worsens the RTT and PLR performance of UEs, regardless of the handover management algorithms. 
This can be attributed to the dynamic link quality and the imperfection of handover timing. 
\textbf{Third,} cell load has a significant impact on the performance of UE. 
Compared Fig.~\ref{fig:ping_RTTandLoss}b against Fig.~\ref{fig:ping_RTTandLoss}a, we see that both RTT and PLR increase dramatically when the cell load becomes heavy.
\textbf{Fourth and most importantly}, \pname outperforms its peer handover strategies in both cases. 
The gain is more significant in a network with heavy traffic load, where \pname has a performance close to the static UE. 
This indicates that the main contribution of RTT and PLR is the resource competition rather than the handover events.


\begin{table}[t]
\centering
\setlength{\tabcolsep}{3pt}
\renewcommand{\arraystretch}{1.1}
\small
\fontsize{7.3}{8.5}\selectfont
\caption{Measured RTT (ms) of a smartphone under different handover policies. `--' indicates that the RTT exceeds the timeout threshold.
M: Method; P: Percentile; L: Light load; H: Heavy load.}
\label{tab:latency_percentile}
\begin{tabular}{c|c|c|c|c|c|c|c|c|c|c|c}
\hline

\multirow{2}{*}{} & M
& \multicolumn{2}{c|}{3GPP \cite{3gpp_sp99100}}
& \multicolumn{2}{c|}{No-MM}
& \multicolumn{2}{c|}{APT \cite{alsuhli2021mobility}}
& \multicolumn{2}{c|}{MR-PHO \cite{hassan2022vivisecting}}
& \multicolumn{2}{c}{\textbf{\pname (ours)}} \\
\cline{2-12}

& P
& 95th & 90th
& 95th & 90th
& 95th & 90th
& 95th & 90th
& 95th & 90th \\
\hline

\multirow{2}{*}{Ping}
& L
& 139 & 57
& 353 & 137
& 121 & 68
& 97 & 54
& \textbf{57} & \textbf{38} \\
\cline{2-12}

& H
& 607 & 283
& -- & 718
& 352 & 201
& 264 & 156
& \textbf{185} & \textbf{118} \\
\hline

\multirow{2}{*}{Iperf}
& L
& 1132 & 400
& 601 & 396
& 356 & 248
& 379 & 303
& \textbf{262} & \textbf{170} \\
\cline{2-12}

& H
& -- & 482
& -- & 451
& 638 & 376
& 470 & 282
& \textbf{318} & \textbf{182} \\
\hline
\end{tabular}
\vspace{-0.2in}
\end{table}

\begin{figure}[t]
    \centering
    \begin{subfigure}[b]{\linewidth}
        \centering
        \includegraphics[width=0.54\linewidth]{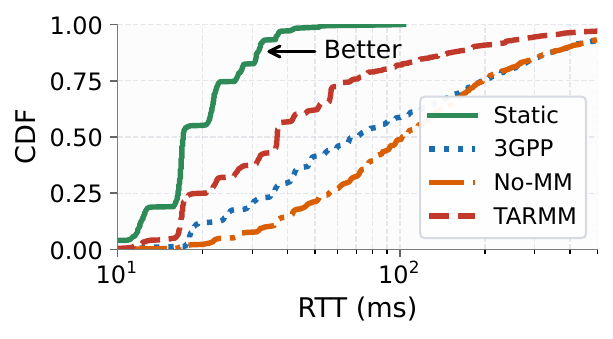}
        \includegraphics[trim=5 -15 8 0, clip, width=0.44\linewidth]{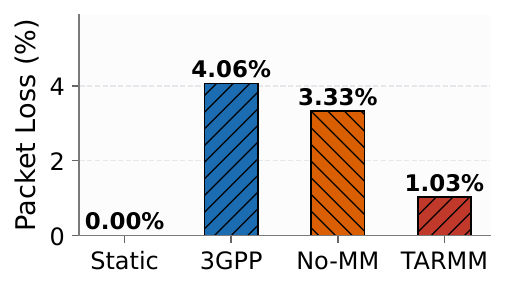}
        \vspace{-0.075in}
        \caption{Cells have light traffic loads.}
        \label{fig:iperf_light_rtt}
    \end{subfigure}
    \hfill
    \begin{subfigure}[b]{\linewidth}
        \centering
        \includegraphics[width=0.54\linewidth]{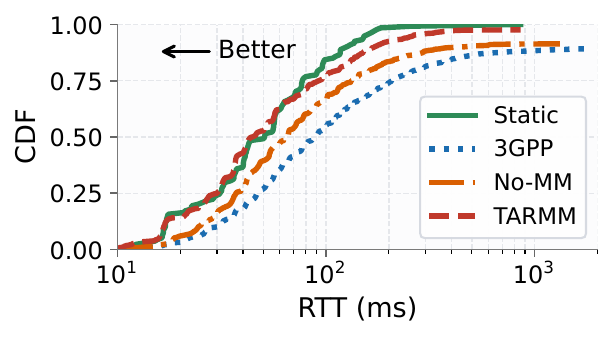}
        \includegraphics[trim=5 -15 8 0, clip, width=0.44\linewidth]{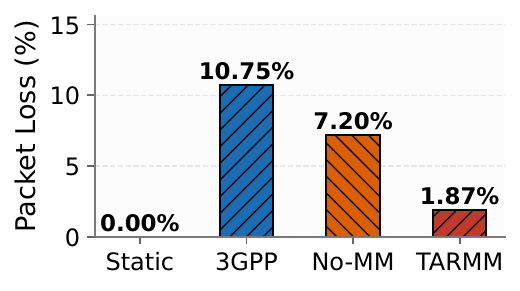}\vspace{-0.075in}
        \caption{Cells have heavy traffic loads.}
        \label{fig:iperf_heavy_rtt}
    \end{subfigure}\vspace{0.075in}    
    \caption{Comprehensive measurement results of RTT and packet loss rate when  UEs \texttt{iperf} an edge GPU server.}\vspace{-0.1in}
    \label{fig:iperf_latency_reliability}
\end{figure}

\textbf{Iperf Data Traffic on UE.}\ \ 
The \texttt{iperf} command is used on the UEs to generate persistent data traffic with large data packets. 
Each of the four UEs generates 1--3~Mbps persistent uplink traffic and 10--30~Mbps persistent downlink traffic, emulating practical edge AI offloading applications.
For this test, we collected 5 hours of data traces in the network over 3 days.
Fig.~\ref{fig:iperf_latency_reliability} presents our experimental results. 
We have the following observations.
\textbf{First}, similar to the previous \texttt{ping} case, \pname demonstrates a significant gain compared to 3GPP and No-MM handover control. 
The gain stems from the fact that \pname considers the global network state encoded by the graph network, rather than using local link information, to make the handover decision for UEs. 
\textbf{Second}, \texttt{iperf} traffic experiences much larger RTT and higher PLR compared to the \texttt{ping} traffic. 
This is expected, because \texttt{iperf} generates more data packets and thus intensifies the resource competition. 
\textbf{Third}, advanced handover management is effective mainly in the networks with low traffic load and becomes ineffective in heavy-loaded network, where resource is the main determining factor other than handover timing. 

\vspace{-0.1in}
\subsection{Comparison with SOTA Learning Methods} 
\label{subsection:State-of-the-Art}

While the previous subsection compared \pname against the existing rule-based handover policies and demonstrated its gain, this subsection focuses on the comparison between \pname and the SOTA learning-based handover methods:  APT \cite{alsuhli2021mobility}
and MR-PHO \cite{hassan2022vivisecting}.
Similar to the previous experiments, we implement APT and MR-PHO in our live network for two smartphones, and the same person carries these three UEs walking around the entire network area. 
The three smartphones generate 3~Mbps uplink and 30~Mbps downlink persistent traffic using \texttt{iperf} command.
In total, 5 hours of data traces have been collected for this comparison.

\begin{figure}[t]
        \centering
        \begin{subfigure}[b]{0.495\linewidth} 
            \centering
            \includegraphics[trim=0 0 0 0, clip, width=\linewidth]{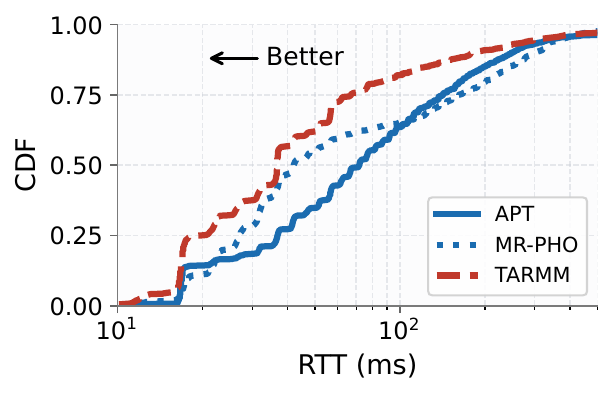}\vspace{-0.075in}
            \caption{Cells have light traffic load.} 
            \label{fig:sota_RTT}
        \end{subfigure}
        \hfill
        \begin{subfigure}[b]{0.495\linewidth}
            \centering
            \includegraphics[trim=0 0 0 0, clip,width=\linewidth]{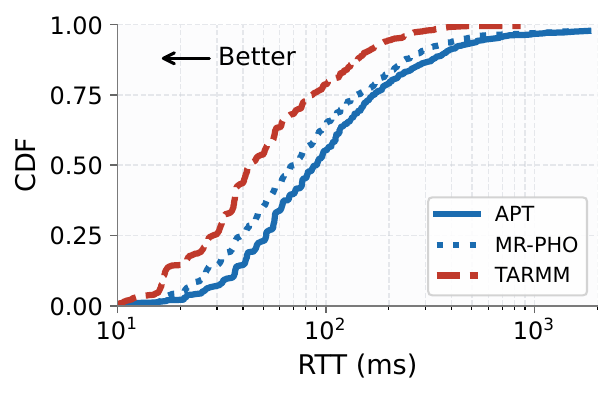}\vspace{-0.075in}
            \caption{Cells have heavy traffic load.} 
            \label{fig:sota_Packet}
        \end{subfigure}\vspace{0.075in}
        \caption{Comparison of \pname against SOTA learning-based handover methods.} \vspace{-0.1in}
        \label{fig:SOTA_comparison}
\end{figure}

Fig.~\ref{fig:SOTA_comparison} reports our experimental results in both light- and heavy-traffic networks. 
\pname consistently outperforms MR-PHO and APT by shifting the RTT distribution left, reducing median latency by about 30--40\% and keeping tail latency (e.g., 90th percentile) significantly lower. It also improves reliability, achieving a lower packet loss rate. 
We attribute the gain of \pname to the temporal graph embedding, making it possible to make handover decision for individual UEs based on global network information rather than local individual link quality. 

Table~\ref{tab:latency_percentile} summarizes the comparison results, demonstrating that \pname outperforms the baselines in mobile scenarios.



\vspace{-0.1in}
\subsection{Ablation Studies}
\label{subsec:Ablation}


We conduct ablation study to evaluate the effectiveness of individual components of \pname. 

\begin{figure}[t]
        \centering
        \begin{subfigure}[b]{0.495\linewidth} 
            \centering
            \includegraphics[trim=0 0 0 0, clip, width=\linewidth]{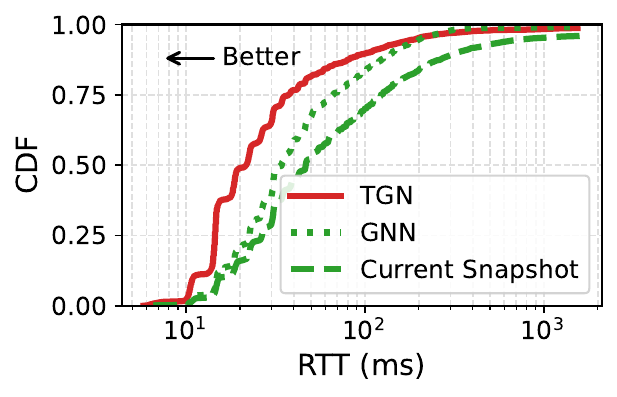}
            \caption{\pname's TGN.} 
            \label{fig:TGNvsGNN}
        \end{subfigure}
        \hfill
        \begin{subfigure}[b]{0.495\linewidth}
            \centering
            \includegraphics[trim=0 0 0 0, clip,width=\linewidth]{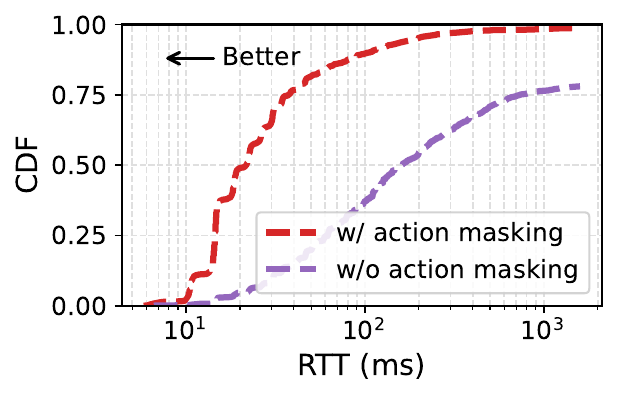}
            \caption{\pname's action masking.}
            \label{fig:masking_impact}
        \end{subfigure}
        \vspace{-0.075in}
        \caption{Experimental results of \pname's ablation studies.} \vspace{-0.1in}
        \label{fig:ablation}
\end{figure}

\textbf{TGN Embedding Module.}\ \
TGN is a key component of \pname, with alternatives including a static GNN and a Current Snapshot approach. While GNN captures spatial relationships through graph aggregation, it ignores temporal dynamics. In contrast, the Current Snapshot approach relies solely on instantaneous local observations for KPM embedding, without incorporating either temporal history or rich spatial context.
Fig.~\ref{fig:TGNvsGNN} compares the performance of \pname when TGN is replaced by these alternatives. TGN significantly shifts the RTT CDF to the left, reducing the median RTT from approximately 40–50\,ms (GNN) and 60–80\,ms (Current Snapshot) to around 20–30\,ms. At the tail, TGN keeps the 90th-percentile RTT below 100\,ms, whereas GNN and Current Snapshot exceed 200\,ms and 400\,ms, respectively.


\begin{figure}[t]
\centering
\includegraphics[trim=0 0 0 0 , clip, width=\linewidth]{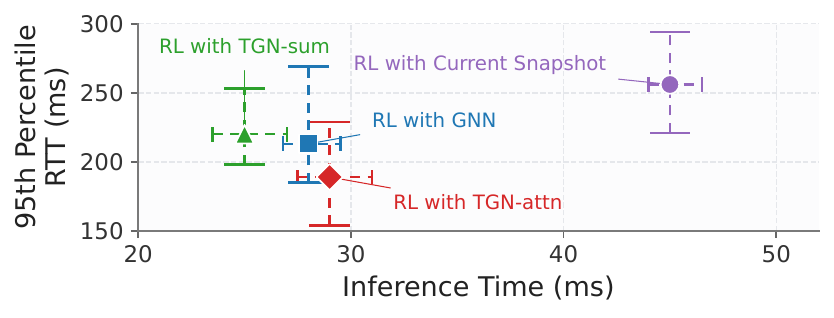}\vspace{0.05in}
\caption{Performance of \pname with different KPM embedding methods.}\vspace{-0.1in}
\label{fig:rl_latency_tradeoff}
\end{figure}

\begin{figure*} [!t]
\centering
\includegraphics[trim=0 0 0 0 , clip, width=1.02\linewidth]{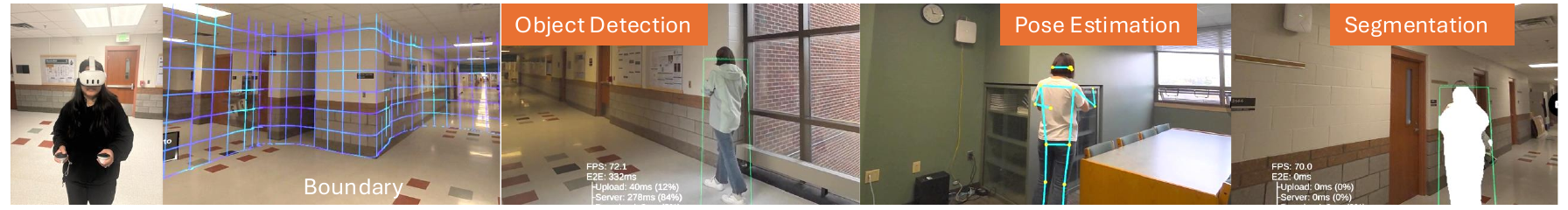}
\caption{Illustration of three VR perception offloading tasks for edge AI computing.}\vspace{-0.15in}
\label{fig:VR_application}
\end{figure*}

\begin{figure}[t]
\centering
\includegraphics[trim=0 0 0 0 , clip, width=\linewidth]{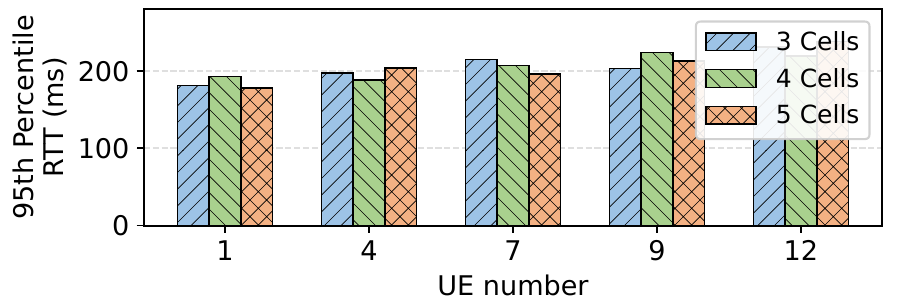}\vspace{0.05in}
\caption{The 95th-percentile RTT when the network has different cells and different number of UEs.
}\vspace{-0.15in}
\label{fig:bar_ue_cell}
\end{figure}

Fig.~\ref{fig:rl_latency_tradeoff} evaluates \pname with different embedding modules in terms of inference time and RTT. Compared to the 3GPP baseline, all RL-based approaches significantly reduce tail latency, though their effectiveness depends on the quality of state representation. RL with a Current Snapshot performs the worst due to the lack of temporal and structural context, while incorporating spatial information via GNN improves performance. TGN-based models further reduce RTT by capturing temporal dynamics. In particular, RL with TGN-attn achieves the best tradeoff, lowering the 95th-percentile RTT to around 200\,ms with minimal inference overhead, highlighting the importance of jointly optimizing model efficiency and network-aware decision-making.

\textbf{Rule-Based Action Masking.}\ \
We compare the performance of \pname with and without the rule-based action masking module. Fig.~\ref{fig:masking_impact} shows that masking significantly improves performance, reducing the median RTT from around 80–100\,ms to 20–30\,ms and lowering the 95th-percentile RTT from over 500\,ms to below 100\,ms. It also eliminates extreme latency events above 1\,s.
These gains stem from constraining unsafe exploration in MARL. Without masking, the agent tends to make aggressive handover decisions for exploration, often selecting transiently strong but unstable links or overloaded cells. Such actions lead to severe performance degradation, including large RTT spikes and increased packet loss.

\vspace{-0.1in}
\subsection{Impacts of Network Size}
\label{subsec:Scalability}
To evaluate the scalability of \pname, we vary the number of active RUs in our 5G O-RAN testbed from 3 to 5, and adjust the number of active UEs in the target cell from 1 to 12.
Fig.~\ref{fig:bar_ue_cell} shows the 95th-percentile RTT measured at the UEs, capturing latency during handover events. The RTT remains relatively stable as the network scales, increasing only modestly from about 180\,ms to 220\,ms. This result demonstrates that \pname is robust to both user density and network size, maintaining stable latency performance under increasing system complexity.


\vspace{-0.1in}
\subsection{VR Offloading Applications}
\label{VR_experiments}

We now extend the evaluation of \pname from a network perspective to application-level performance. We consider three YOLO-based VR perception tasks on Meta Quest 3: \textit{Object Detection}, \textit{Pose Estimation}, and \textit{Semantic Segmentation}, as shown in Fig.~\ref{fig:VR_application}. 
For each task, the VR headset captures images, compresses them, and offloads them to an edge GPU server over the 5G network. The GPU server performs inference and returns the results for real-time rendering. 
The image frame rate is set to 20 FPS. Since Quest 3 does not support 5G connectivity, a smartphone hotspot is used to bridge the connection to the 5G network.

\begin{figure}[t]
        \centering
        \begin{subfigure}[b]{0.48\linewidth} 
            \centering
            \includegraphics[trim=0 0 0 0, clip, width=1.05\linewidth]{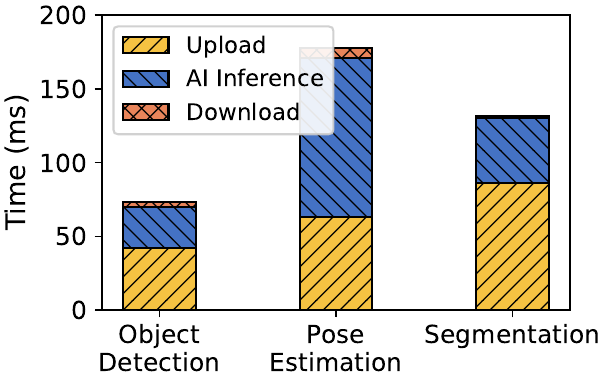}
            \caption{Latency Breakdown.} 
            \label{fig:stacked_bar_latency}
        \end{subfigure}
        \hfill
        \begin{subfigure}[b]{0.51\linewidth}
            \centering
            \includegraphics[trim=0 0 0 0, clip,width=\linewidth]{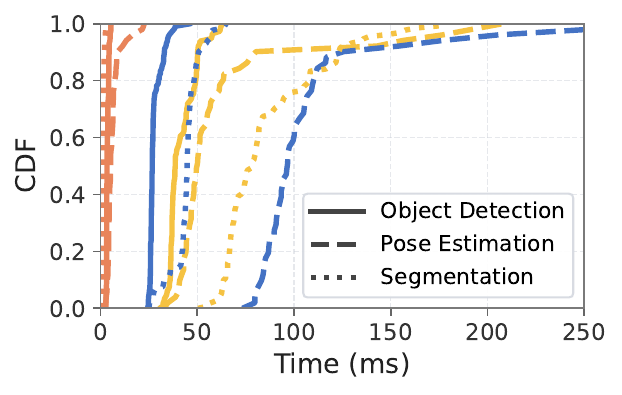}
            \caption{CDF of breakdown latency.}
            \label{fig:stacked_bar_latency_cdf}
        \end{subfigure}
        \caption{Average and CDF of delay breakdown components under different VR offloading tasks. [Yellow: upload time; blue: inference time; orange: download time.]} \vspace{-0.1in}
        \label{fig:Latency_Breakdown}
\end{figure}

\textbf{End-to-End Delay Profile.}
The three applications exhibit distinct data transmission characteristics. Object detection and pose estimation generate approximately 22\,KB uplink and 2\,KB downlink per frame, while Semantic Segmentation produces significantly larger traffic (110\,KB uplink and 6\,KB downlink).
Fig.~\ref{fig:Latency_Breakdown} shows the breakdown of end-to-end delay across workloads. As illustrated in Fig.~\ref{fig:stacked_bar_latency}, object detection achieves the lowest overall latency, pose estimation is dominated by inference delay, and semantic segmentation incurs the highest upload latency due to larger input size. This highlights heterogeneous bottlenecks across workloads. Fig.~\ref{fig:stacked_bar_latency_cdf} further presents the latency distributions: object detection exhibits consistently low latency with a tight distribution, pose estimation shows moderate tail latency, while semantic segmentation has significantly heavier tails, with the 95th-percentile exceeding 100\,ms. These results indicate that compute- and bandwidth-intensive workloads are more sensitive to network dynamics and resource allocation.

\begin{figure}[t]
    \centering
    \begin{subfigure}[b]{0.325\linewidth} 
        \centering
        \includegraphics[width=1.06\linewidth]{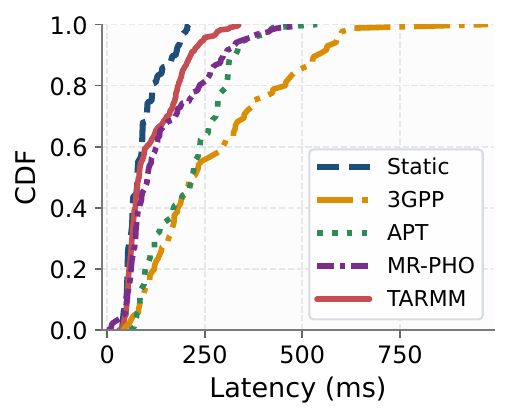} 
        \caption{Object detection.} 
        \label{fig:first_sub}
    \end{subfigure}
    \hfill
    \begin{subfigure}[b]{0.325\linewidth}
        \centering
        \includegraphics[width=1.06\linewidth]{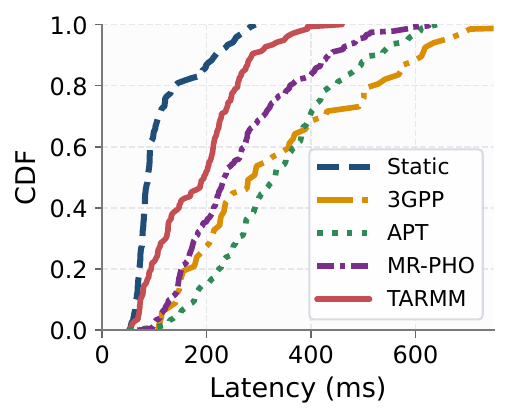}
        \caption{Pose estimation.}
        \label{fig:    }
    \end{subfigure}
    \hfill
    \begin{subfigure}[b]{0.325\linewidth}
        \centering
        \includegraphics[width=1.06\linewidth]{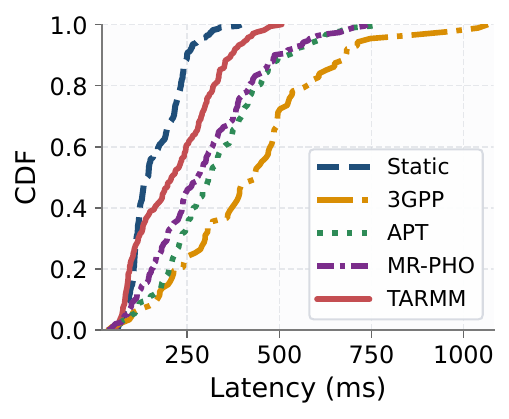}
        \caption{Segmentation.}
        \label{fig:}
    \end{subfigure}
    
    \caption{End-to-end delay comparison of different mobility management schemes for different edge AI tasks.}\vspace{-0.1in}
    \label{fig:SOTA_VR}
\end{figure}

\textbf{Comparison with SOTA Methods.}
We compare \pname with rule-based and learning-based handover methods in terms of end-to-end delay, including upload, inference, and download time consumptions. Fig.~\ref{fig:SOTA_VR} shows that \pname consistently outperforms all mobility management baselines across the three applications, with larger gains for latency-sensitive workloads. Specifically, \pname reduces median latency by 30–40\% and 90th-percentile latency by about 35\% compared to the 3GPP baseline, with the most significant improvement observed for segmentation. Compared to learning-based baselines (APT and MR-PHO), \pname achieves an additional 15–25\% reduction in both median and tail latency. These results demonstrate that \pname effectively improves both typical and worst-case delay by enabling timely and load-aware handover decisions across diverse edge AI workloads.


\textbf{Quantitative User Experience.}
Since VR user experience is inherently subjective, we quantify it using two metrics measured on VR headsets.
\textit{(i) Frame Drop Rate}, which captures VR headset's rendering continuity under network dynamics, and
\textit{(ii) Frame Freeze Ratio}, which measures the fraction of time the display stalls due to delayed or missing frames, directly reflecting perceived smoothness.
These two metrics are directly reflective of user experience \cite{van2022quality, tsui2026fovrl, tripathi2024assessing}.

\begin{figure}[t]
\centering
\includegraphics[trim=0 0 0 0 , clip, width=\linewidth]{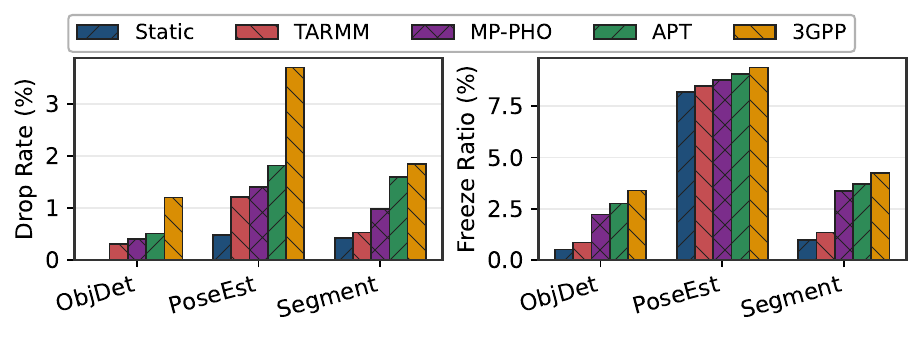}\vspace{0.05in}
\caption{VR's user experience comparison of  different mobility management schemes.}\vspace{-0.1in}
\label{fig:User_Experience}
\end{figure}

Fig.~\ref{fig:User_Experience} presents the measurement results for three representative VR offloading tasks under different handover schemes. \pname consistently achieves lower frame drop rates and smaller frame freeze ratios than all baselines, with the largest gains observed in throughput-demanding workloads such as PoseEst and Segment, where mobility-induced latency variations are more disruptive. In contrast, the rule-based 3GPP baseline exhibits the worst performance due to reactive, signal-only decisions that incur higher packet loss and queuing delay. Learning-based methods (MP-PHO and APT) offer moderate improvements over 3GPP but still lag behind \pname, as they do not effectively capture temporal dynamics or network conditions. These results demonstrate that reducing handover-induced disruptions directly translates into improved application-level QoE, as well as the effectiveness of \pname from an application perspective.

\vspace{-0.1in}
\section{Related Work}




\vspace{-0.1in}
\textbf{Latency Optimization in 5G.}
Reducing end-to-end latency has been a central objective in 5G networks, particularly for Ultra-Reliable Low-Latency Communications (URLLC) and delay-sensitive applications. 
Prior work has explored latency optimization from multiple perspectives, including physical-layer enhancements \cite{samuylov2025empowering}, scheduling and resource allocation \cite{kim2024distributed,wang2023low}, and transport/network-layer optimizations such as edge deployment and protocol tuning \cite{abkenar2022survey, langley2017quic, ahmad2025warping}. 
In addition, several studies have proposed cross-layer designs that jointly consider radio conditions, traffic dynamics, and queueing behavior to minimize latency \cite{tan2021device, xing2025rajomon, shenmortise, shen2026law}. 

\textbf{Handover Management.}
Handover management has been extensively studied with various objectives, including maximizing throughput \cite{mehregan2025gcn}, improving load balancing \cite{chang2022decentralized,pateromichelakis2014graph}, enhancing fairness \cite{prado2023enabling}, and reducing signaling overhead \cite{shen2025decentralized, tan2021device, tayyab2019signaling}. 
A wide range of techniques have been proposed, such as rule-based heuristics \cite{tayyab2019survey}, optimization-based formulations \cite{mollel2021survey}, and learning-based approaches such as reinforcement learning \cite{alsuhli2021mobility,huang2020closer} and predictive models \cite{wang2023mitigating,li2020beyond}. These methods aim to improve network efficiency by adapting handover thresholds, predicting mobility patterns, or optimizing cell selection policies.

However, most prior work relies on simplified mathematical models or simulated datasets that fail to capture the complexity and dynamics of real-world networks.
Moreover, most studies were evaluated primarily through numerical analysis, simulation, or network emulation, which often do not reflect the operational constraints and variability of deployed systems. 
\pname advances this research line by developing mobility management from an O-RAN system perspective and evaluating it on a realistic O-RAN testbed.

\textbf{Learning-based Network Control.}
Learning-based approaches have been widely adopted as a general framework for O-RAN testing \cite{tan2025automated}, network control\cite{hurtado2022deep, maleki2024qos}, improving QoE \cite{ye2024dissecting, yan2025near, lu2026eexapp},enabling adaptive decision-making\cite{boutaba2018comprehensive} in complex and dynamic environments. 
A wide range of techniques have been proposed, including supervised learning for traffic and performance prediction \cite{perry2023dote}, optimization-guided learning frameworks \cite{mao2017neural}, and reinforcement learning for sequential decision-making \cite{yan2025xdiff}. 
More recently, graph-based models have been introduced to capture spatial dependencies among network entities \cite{xu2023teal}, while multi-agent learning frameworks enable decentralized control across distributed network components \cite{mao2019learning}. 
These approaches aim to learn adaptive control policies directly from data and interactions with the environment.
\pname differs from them in both objective and approach.

\vspace{-0.1in}
\section{Conclusion}
\vspace{-0.1in}
In this paper, we presented \pname, a 5G O-RAN system that optimizes user mobility management for delay-critical edge AI offloading. By leveraging temporal graph embedding, \pname captures the spatiotemporal dynamics of user mobility and network conditions, enabling more informed and proactive handover decisions. To ensure safe and deployable operation, we incorporate rule-based action masking, which constrains MARL exploration using domain knowledge and prevents harmful or infeasible actions. We implement \pname on a real-world testbed and demonstrate its effectiveness under both network-level and application-level scenarios. Overall, this work highlights the importance of integrating learning-based decision-making with domain knowledge and system-level design in future RANs.


\clearpage

\bibliographystyle{plain}
\bibliography{8_references}


\clearpage

\appendix
\section*{Appendix}
\section{Actor-Critic Loss}\label{app:ppo}

The actor and critic are optimized using PPO with a clipped surrogate objective to ensure stable policy updates.

\begin{align}
\mathcal{L}_{\text{actor}} 
&= -\mathbb{E}_t \Big[
\min\big(
r_t(\theta)\hat{A}_t,\;
\tilde{r}_t(\theta)\hat{A}_t
\big)
\Big]
- c_H \mathcal{H}(\pi_\theta), \\
\tilde{r}_t(\theta)
&= \mathrm{clip}(r_t(\theta), 1-\epsilon, 1+\epsilon) \notag \\
\mathcal{L}_{\text{critic}} 
&= \mathbb{E}_t \Big[
\max\big(
(V_\phi(o_t) - R_t)^2,\;
(\tilde{V}_t - R_t)^2
\big)
\Big], \\
\tilde{V}_t
&= \mathrm{clip}(V_\phi(o_t), 
V_{\phi_{\mathrm{old}}} - \epsilon,\;
V_{\phi_{\mathrm{old}}} + \epsilon) \notag
\end{align}
where $r_t(\theta) = \frac{\pi_\theta(a_t \mid o_t)}{\pi_{\theta_{\mathrm{old}}}(a_t \mid o_t)}$ 
denotes the importance sampling ratio, and $\hat{A}_t$ is the advantage estimate, computed using generalized advantage estimation (GAE). 
The clipping operation restricts policy updates within a trust region to prevent excessively large policy shifts. 
$c_H$ is the entropy coefficient that encourages exploration.
For the critic, $V_\phi(o_t)$ is the value function and $R_t$ is the empirical return. 
The clipped value loss stabilizes training by limiting large deviations from the previous value estimate $V_{\phi_{\mathrm{old}}}(o_t)$.
The policy $\pi_\theta(a_i \mid \mathbf{o}_i)$ takes as input the TGN embedding $\mathbf{o}_i = \mathbf{z}_i(t)$, and the value function $V_\phi(\mathbf{o}_i)$ takes as input the concatenation $\mathbf{s}_i = [\mathbf{z}_i(t), \mathbf{g}(t)]$, both parameterized by $\psi$.

\section{Embedding Interpretability}\label{app:embedding}

\begin{figure}[h]
        \centering
        \begin{subfigure}[b]{0.9\linewidth} 
            \centering
            \includegraphics[trim=0 0 -30 0, clip, width=\linewidth]{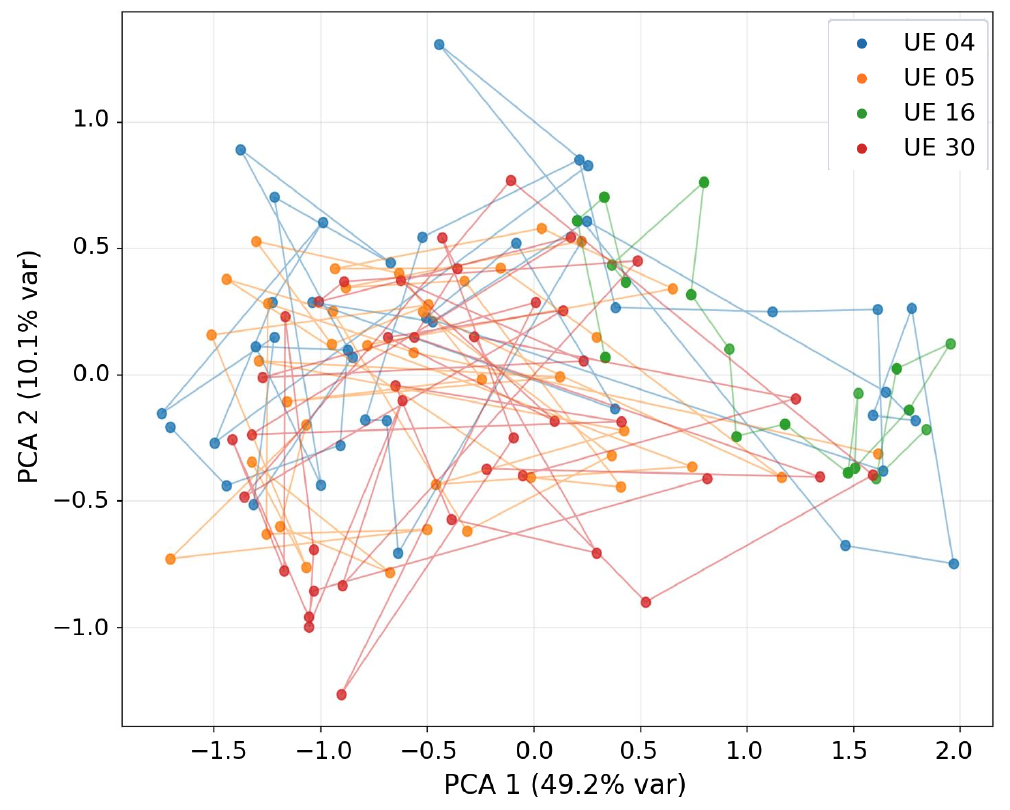}
            \caption{PCA Visualization of UE Embedding.} 
            \label{fig:tgn_pca}
        \end{subfigure}
        \hfill
        \begin{subfigure}[b]{0.95\linewidth}
            \centering
            \includegraphics[trim=0 0 0 0, clip,width=\linewidth]{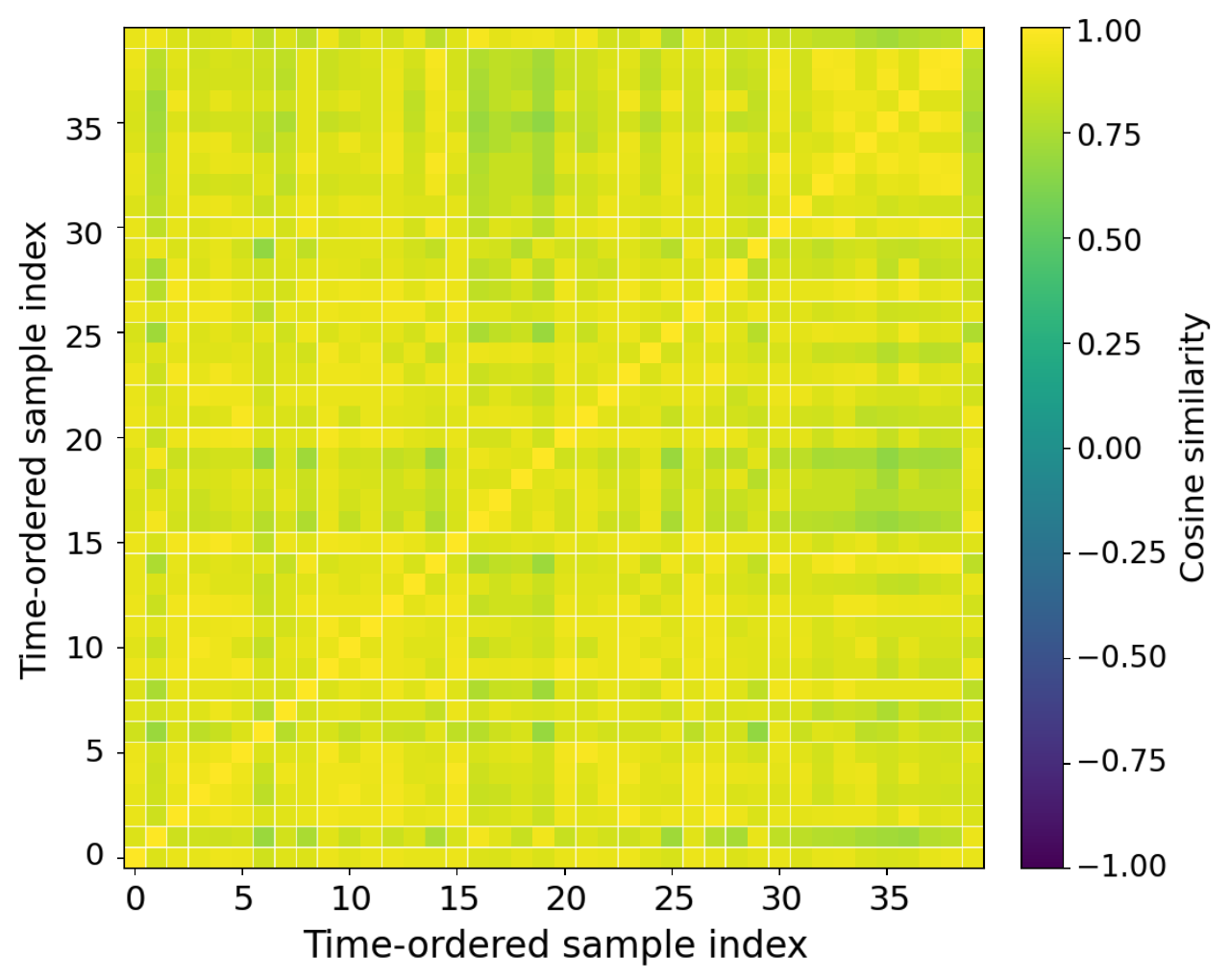}
            \caption{Pairwise Cosine Similarity of Time-Ordered UE Embeddings.}
            \label{fig:tgn_representation}
        \end{subfigure}
        \caption{Temporal Dynamics and Consistency of TGN-Generated UE Embeddings} 
        \label{fig:tgn_repre}
\end{figure}

To better understand the learned temporal representations, we visualize the UE embeddings generated by TGN.
Fig.~\ref{fig:tgn_pca} shows a PCA projection of UE embeddings over time for multiple UEs. 
Each trajectory corresponds to a single UE, where consecutive points are connected in temporal order. 
We observe that embeddings evolve smoothly over time while remaining distinguishable across UEs, indicating that the TGN encoder captures both temporal continuity and UE-specific characteristics.
Fig.~\ref{fig:tgn_representation} presents the pairwise cosine similarity matrix of time-ordered embeddings. 
The strong diagonal structure reflects high similarity between temporally adjacent samples, 
confirming temporal consistency in the learned representations. 
Meanwhile, the absence of uniformly high similarity across the matrix suggests that the embeddings preserve sufficient diversity to distinguish different network states.

\section{Delay Requirements}

Delay-critical AI applications are increasingly deployed in domains such as real-time video analytics, AR/VR, autonomous robotics (e.g., remote surgery), and industrial automation \cite{Nokia2022XR, AssemblyAI}. 
In these applications, timely data delivery is critical to ensure responsive and reliable system behavior. 
AI inference often operates in closed-loop pipelines in which sensor data must be transmitted to edge or cloud servers for processing, and the resulting decisions\cite{raffik2025edge} must be returned within strict latency constraints. 
Table~\ref{tab:latency_ai} shows the end-to-end delay requirements of edge AI offloading for some emerging applications, which includes both (uplink and downlink) communications delays as well as edge computing delays.
The string delay requirements pose a grand challenge in the design and optimization of wireless networks. 

\begin{table}
\fontsize{7}{11}\selectfont
\caption{End-to-end latency requirements for some edge AI offloading applications.}
\label{tab:latency_ai}
\centering
\resizebox{\linewidth}{!}{
\begin{tabular}{@{}p{1.6cm}l@{\hspace{-2pt}}c@{}}
\hline
\textbf{Category} & \textbf{Edge AI Offloading Tasks} & \textbf{Latency Requirement} \\
\hline

\multirow{2}{*}{Immersive AI} 
& VR/AR Motion-to-Photon\cite{8329628,fornes2025acceptable} & AR$<$10ms VR$<$20 ms \\
& Cloud XR rendering\cite{nguyen2020low} & $<$50 ms \\

\hline
\multirow{2}{*}{Haptic AI} 
& Remote Robotic Surgery\cite{motiwala2025telesurgery} & Vision$<$150 ms, Touch$<$10 ms \\
& Industrial tele-robotics\cite{cardinaels2025challenges} & $<$10 ms \\


\hline
\multirow{2}{*}{Industrial Edge AI} 
& Collaborative Robots\cite{kang2024robot} & $<$20 ms \\
& AI Vision Inspection \cite{raffik2025edge} & $<$50 ms \\

\hline
\multirow{1}{*}{General AI} 
& High-Frequency Trading\cite{luxalgo_latency_trading} & $<$100 ms \\

\hline
\end{tabular}}\vspace{-0.2in}
\end{table}

\end{document}